\documentclass[reprint,superscriptaddress,preprintnumbers,amsmath,amssymb,aps,prd,tightenlines,longbibliography,balancelastpage]{revtex4-2}

\usepackage{graphicx}
\usepackage[dvipsnames]{xcolor}
\usepackage[sort&compress]{natbib}
\usepackage{amsmath,amssymb,bm,bbm,slashed,subdepth}
\usepackage{xr-hyper}
\usepackage[unicode=true,pdfusetitle,
 bookmarks=true,bookmarksnumbered=false,bookmarksopen=false,
 breaklinks=false,pdfborder={0 0 1},backref=false,colorlinks=true]
 {hyperref}
\hypersetup{
 pdfauthor={Grant N. Remmen and Nicholas L. Rodd},
colorlinks=true
,urlcolor=blue
,anchorcolor=blue
,citecolor=blue
,filecolor=blue
,linkcolor=red
,menucolor=blue
,linktocpage=true
,pdfproducer=medialab
,pdfa=true}
\usepackage{cleveref}
\usepackage{enumerate}
\usepackage{epsfig, subfigure}
\usepackage{setspace}
\usepackage{booktabs, tabularx}
\usepackage{units}
\usepackage{placeins}
\usepackage{multirow}
\usepackage{mathtools}
\usepackage[normalem]{ulem}
\usepackage{parskip}
\newcommand{\be}{\begin{equation}}
\newcommand{\ee}{\end{equation}}
\newcommand{\bea}{\begin{equation}\begin{aligned}}
\newcommand{\eea}{\end{aligned}\end{equation}}

\newcommand{\Fig}[1]{Fig.~\ref{#1}}
\newcommand{\Eq}[1]{Eq.~\eqref{#1}}
\newcommand{\mysec}[1]{\noindent {\bf #1.}}

\usepackage{ragged2e}
\usepackage{etoolbox}
\apptocmd{\thebibliography}{\justifying\setlength{\leftskip}{7.4mm}}{}{} 

\newcommand{\OO}{\mathcal{O}}

\newcommand{\OSa}{\OO_+}
\newcommand{\cSa}{C_+}
\newcommand{\OSb}{\OO_\times}
\newcommand{\cSb}{C_\times}
\newcommand{\OSc}{\OO_-}
\newcommand{\cSc}{C_-}

\newcommand{\OHa}{\OO_1^{h}}
\newcommand{\cHa}{c_1^{h}}
\newcommand{\OHb}{\OO_1^{h\pi}}
\newcommand{\cHb}{c_1^{h\pi}}
\newcommand{\OHc}{\OO_2^{h\pi}}
\newcommand{\cHc}{c_2^{h\pi}}
\newcommand{\OHd}{\OO_1^{\pi}}
\newcommand{\cHd}{c_1^{\pi}}
\newcommand{\OHe}{\OO_2^{\pi}}
\newcommand{\cHe}{c_2^{\pi}}

\begin{document}

\title{Positively Identifying HEFT or SMEFT}

\author{Grant N. Remmen}
\affiliation{Center for Cosmology and Particle Physics, Department of Physics, New York University, New York, NY 10003}
\author{Nicholas L. Rodd}
\affiliation{Theory Group, Lawrence Berkeley National Laboratory, Berkeley, CA 94720}
\affiliation{Berkeley Center for Theoretical Physics, University of California, Berkeley, CA 94720}

\begin{abstract}
\noindent 
We establish the bounds on Wilson coefficients of the Higgs effective field theory (HEFT) mandated by unitarity and analyticity.
These positivity constraints can be projected into the space of the standard model effective field theory (SMEFT) as HEFT$\,\supset\,$SMEFT.
Doing so reveals a subspace allowed by the HEFT but forbidden by SMEFT positivity, thereby identifying a region that could herald the use of the wrong EFT rather than a pathological UV.
Restricting to custodial symmetric dimension-eight Higgs operators, there is a unique pair within the SMEFT where this concept can be sharply realized and is already being probed at colliders.
\end{abstract}
\maketitle

\begin{figure}[t]
\begin{center}
\includegraphics[width=7.65cm]{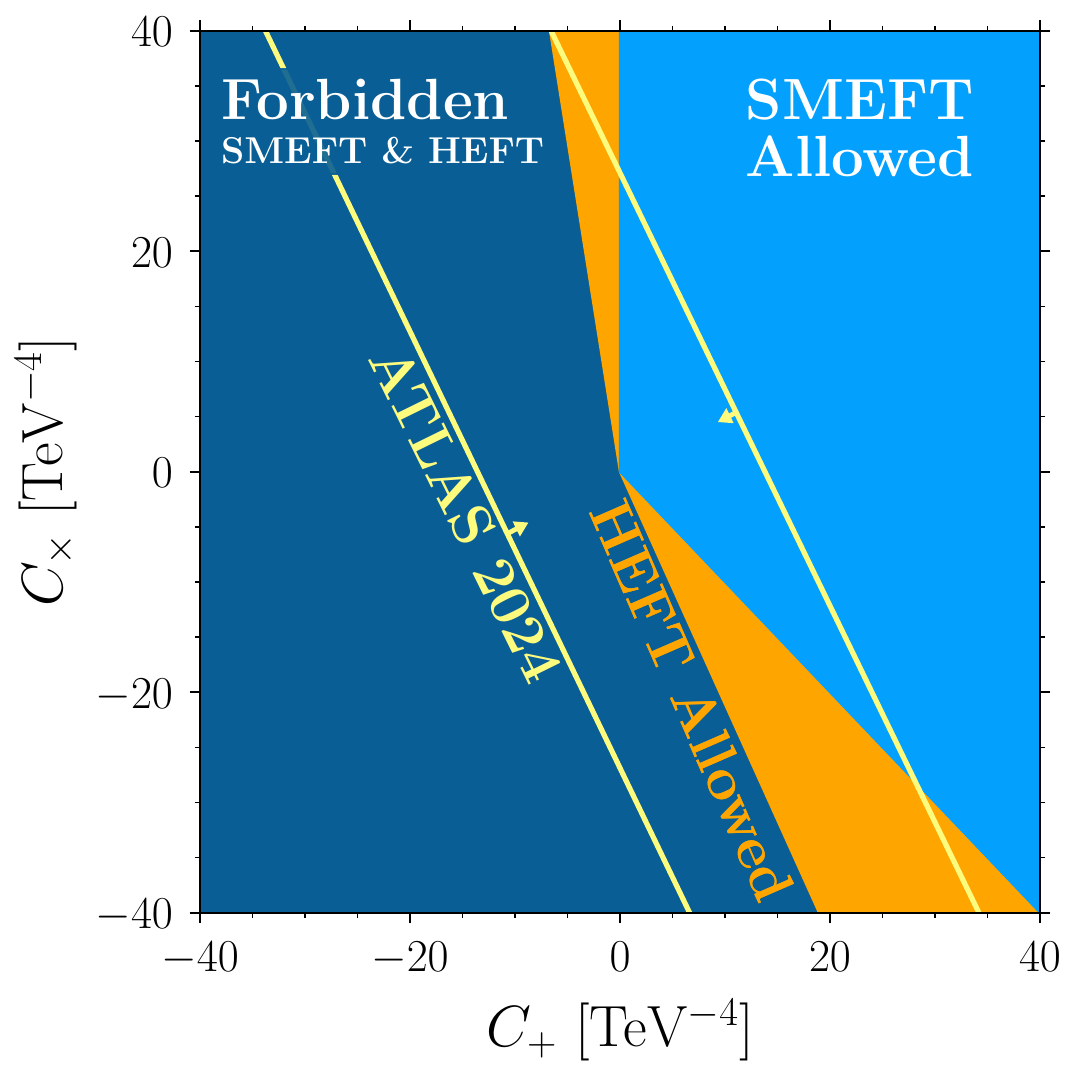}
\end{center}\vspace{-4mm}
\caption{Positivity bounds for the HEFT projected into the two-dimensional space spanned by the custodial symmetric SMEFT.
Applying positivity to the SMEFT directly restricts the allowable EFTs to live in the light blue region.
If the Higgs sector is instead described by the HEFT, this region is enlarged to include the region shown in orange: a unique region within which new physics could emerge as indicating a breakdown of the SMEFT rather than violation of analyticity or unitarity.
ATLAS measurements constrain these coefficients to within the yellow lines~\cite{ATLAS:2023sua}.
We justify these partitions in the present Letter.\vspace{-5mm}}
\label{fig:proj}
\end{figure}

%%%%%%%%%%%%%%%%%%%%%%%%%%%%%%%%%
\mysec{Introduction}
%%%%%%%%%%%%%%%%%%%%%%%%%%%%%%%%%
%
In the absence of a discovery of novel particles at the energy frontier, the possibility has sharpened that new physics may first appear indirectly through precision tests of the standard model (SM).
But how can we characterize the impact of states that exist at scales our experiments cannot yet reach?
The answer is provided by effective field theory (EFT), a broad set of tools developed in the middle decades of the twentieth century through which the effects of new physics can be parameterized even before it is discovered.
With EFT, the influence of ultraviolet (UV) states is consistently encoded in operators of higher mass dimension that modify the interactions of the particles accessible at low energies.

Remarkably, the full space of couplings of these higher-dimension operators---dubbed Wilson coefficients---is not completely spanned by healthy UV theories.
If we assume the UV respects the axioms of our most successful theories---unitarity, locality, and causality---then its shadow in the IR only allows couplings with certain signs and magnitudes~\cite{Adams:2006sv,Jenkins:2006ia,Dvali:2012zc,Nicolis:2009qm,Arkani-Hamed:2020blm,Bellazzini:2020cot}.
Such positivity bounds have found application to a stunning array of EFTs~\footnote{For instance, among many possible examples ranging from phenomenology to formal theory, positivity has been applied to quantum gravitational modifications of general relativity and their implications for the Weak Gravity Conjecture~\cite{Bellazzini:2015cra,Cheung:2016wjt,Camanho:2014apa,Gruzinov:2006ie,Arkani-Hamed:2021ajd,Cheung:2014ega,Bellazzini:2019xts,Andriolo:2020lul,Caron-Huot:2021rmr,Caron-Huot:2022ugt,Caron-Huot:2022jli,Cheung:2016yqr,deRham:2017xox,Camanho:2016opx}.}, the most important of which for our present discussion is that of the SM (SMEFT)~\cite{Remmen:2019cyz,Remmen:2020vts,Remmen:2022orj,Bi:2019phv,Zhang:2018shp,Remmen:2020uze,Low:2009di,Englert:2019zmt,ZZ,YZZ,Trott,HYZZ,Chala:2023xjy,Hong:2024fbl,Chen:2023bhu}.
This success begs the question: If a detection is made of a Wilson coefficient violating positivity bounds, what does this mean?
The apparent implications would be profound, suggesting the breakdown of a basic field theory axiom at energy scales near experimental reach.
However, an apparent violation could also be generated by the use of the wrong EFT.

An EFT is constructed from both the low-energy field content and assumed symmetries.
However, which symmetries to infer from the {\it relevant}  (mass dimension four and lower) operators in constructing the {\it irrelevant} (mass dimension five or greater) ones can be ambiguous.
In the SM, a natural choice is ${\rm SU}(3)_C \times {\rm SU}(2)_L \times {\rm U}(1)_Y$, and imposing this symmetry generates the SMEFT.
Below the electroweak scale, however, what is realized is ${\rm SU}(3)_C \times {\rm U}(1)_{\rm EM}$, and imposing this less restrictive symmetry gives rise to the Higgs EFT (HEFT).
That is, in the HEFT, electroweak symmetry is nonlinearly realized; while this represents a mere field redefinition in the SM, leaving the S-matrix invariant, it has significant consequences for the EFT, as the four real scalars associated with the Higgs field $H$ are no longer required to transform together as an electroweak doublet.
Although an old topic~\cite{Feruglio:1992wf,Bagger:1993zf,Koulovassilopoulos:1993pw}, the HEFT is being actively developed: from its operator counting~\cite{Alonso:2012px,Brivio:2013pma,Graf:2022rco}, geometric interpretation~\cite{Cohen:2021ucp,Alonso:2015fsp,Alonso:2016oah}, and phenomenological necessity~\cite{Alonso:2015fsp,Alonso:2016oah,Falkowski:2019tft,Cohen:2020xca,Banta:2021dek,Gomez-Ambrosio:2022why}.
(For a more comprehensive review of the HEFT, see Refs.~\cite{Brivio:2016fzo,Brivio:2017vri,Alonso:2023upf}.)

What have not been carefully studied are positivity bounds within the HEFT.
In particular, how should one interpret SMEFT positivity bounds if the UV in fact generates the HEFT?
These issues are the focus of this work, and an answer is provided already in \Fig{fig:proj}.
There, we depict the region consistent with SMEFT positivity in the parameter space of the two custodial symmetric dimension-eight Higgs operators in the theory, both of the parametric form $(\partial H)^4$, which we define in detail shortly.
We further show how this region is extended from the application of positivity to the HEFT.
In this Letter, we construct the pertinent HEFT operators---there are five---and derive the associated positivity bounds resulting from analyticity and unitarity.
We construct these bounds using both elastic forward scattering and the generalized optical theorem techniques of Refs.~\cite{ZZ,YZZ,Trott,Arkani-Hamed:2021ajd,HYZZ,Freytsis:2022aho}.
Building example UV completions, we find that the full allowed space of Wilson coefficients can be spanned by simple one-particle extensions, so these operators are of realistic phenomenological interest.
These HEFT bounds are then projected down into the SMEFT subspace, obtaining the region shown in \Fig{fig:proj} that would indicate new physics associated with the HEFT rather than a perverse UV.
Although the HEFT and SMEFT subspaces are not disjoint, if the unique HEFT region were not included, discoveries could be erroneously associated with a violation of positivity.

Our observation is not purely esoteric: the depicted parameter space is already experimentally accessible through collider searches for anomalous quartic gauge couplings (aQGCs)~\cite{Eboli:2006wa}---the complete basis of which was recently constructed in Ref.~\cite{Durieux:2024zrg}---and is being actively constrained at the Large Hadron Collider.
Indeed, \Fig{fig:proj} includes ATLAS constraints derived from a search for final states involving a pair of same-sign $W$ bosons and two jets~\cite{ATLAS:2023sua}.
Assuming a linear representation of the SM gauge symmetry (i.e., SMEFT), the ATLAS measurements constrain certain $(DH)^4$ operators, which we detail shortly; moreover, if one instead assumes a nonlinear realization (i.e., HEFT) and then projects down to the SMEFT subspace, regardless of the UV physics~\footnote{We note that it is never possible for any combination of SMEFT-like and HEFT-like UV physics to produce Wilson coefficients that, upon projection to the SMEFT, land in the forbidden dark blue region in \Fig{fig:proj}. This is clear since every UV completion that transforms linearly under the SM gauge group (producing the SMEFT) can be written as a special case of a HEFT completion.}, one finds that the various regions allowed or forbidden in HEFT or SMEFT---detailed in our analysis below---intersects nontrivially with the space permitted by the ATLAS measurement.

%%%%%%%%%%%%%%%%%%%%%%%%%%%%%%%%%
\medskip
\mysec{Custodial SMEFT and HEFT}
%%%%%%%%%%%%%%%%%%%%%%%%%%%%%%%%%
%
We begin by establishing the operator basis.
In order to maximize any distinction between the SMEFT and HEFT, we focus on the Higgs sectors of both theories.
Further, as reviewed below, sharp positivity bounds arise most directly from dimension-eight operators that support two-to-two scattering amplitudes.
Accordingly, the operators of interest take the schematic form $(DH)^4$.
Operators at this mass dimension but lower order in derivatives, such as $X H^2 (DH)^2$, are higher order in field insertions and therefore do not contribute to quartic Higgs scattering at leading order in the EFT coefficients, which is our focus.

Within the SMEFT, there are three operators of this type~\cite{Hays:2018zze,Eboli:2006wa}.
For our purposes, it is convenient to write these operators as follows,
\bea
{\cal L} &\supset \cSa \OSa + \cSc \OSc + \cSb \OSb
\\
\OSa &= (\partial_{(\mu} H^\dagger \partial_{\nu)} H)(\partial^{(\mu} H^\dagger \partial^{\nu)} H) 
\\
\OSc &= (\partial_{[\mu} H^\dagger \partial_{\nu]} H)(\partial^{[\mu} H^\dagger \partial^{\nu]} H)
\\
\OSb &= (\partial^\mu  H^\dagger \partial_\mu H)(\partial^\nu H^\dagger \partial_\nu H).
\label{eq:SMEFT}
\eea
Regarding notation, we write $D_\mu \rightarrow \partial_\mu$, as the distinction is irrelevant in constructing positivity bounds. Higgs SU(2)$_L$ indices are contracted among terms within parentheses.
For Lorentz indices, we use round or square brackets to denote normalized symmetrization or antisymmetrization, respectively, i.e., $T_{(\mu\nu)}=(T_{\mu\nu}+T_{\nu\mu})/2$.

To isolate the essential physics, we demand an additional symmetry of the UV: custodial invariance.
Custodial symmetry is the O(4) invariance of the SM Higgs sector as one transforms among the four scalar degrees of freedom in $H$; after electroweak symmetry breaking, the symmetry is spontaneously broken down to O(3) by the Higgs vacuum expectation value (vev).
Imposing custodial invariance mandates $\cSc=0$ and thereby reduces the SMEFT to the following space of two operators,
\be
{\cal L} \supset \cSa  \OSa + \cSb \OSb.
\label{eq:SMEFTcustodial}
\ee
The coefficients of these operators form the axes in \Fig{fig:proj}.

Turning to the HEFT, in general, there need be no relation among the four scalar degrees of freedom that combine to form $H$ in the SMEFT.
There are then 55 dimension-eight four-scalar operators to consider~\footnote{For a collection of $n$ scalars $\Phi_I$, there are $n(n\,{+}\,1)(n^2\,{+}\,n\,{+}\,2)/8$ EFT operators of the form $c_{IJKL}\partial_\mu \Phi_I \partial^\mu \Phi_J \partial_\nu \Phi_K \partial^\nu \Phi_L$.
The CP-violating operator $\epsilon^{\mu \nu \rho \sigma} \partial_{\mu} \Phi_{I} \partial_{\nu} \Phi_J \partial_{\rho} \Phi_K \partial_{\sigma} \Phi_{K}$ can be written as a total derivative upon integration by parts.}.
Custodial invariance provides a dramatic simplification.
Following Ref.~\cite{Alonso:2016oah}, we decompose $H$ into four scalars $h$ and $\pi_i$ that transform as a singlet and fundamental under custodial O(3) symmetry.
With this restriction, explicit calculation reveals five remaining independent HEFT operators, up to total derivatives and field redefinitions,
\bea
{\cal L} &\supset \cHa \OHa + \cHb \OHb +\cHc \OHc +\cHd \OHd +\cHe \OHe 
\\
\OHa &=(\partial h)^{4}
\\
\OHb &=(\partial h)^{2}(\partial_{\mu}\pi_{i}\partial^{\mu}\pi_{i})
\\
\OHc &=(\partial_{\mu}h\partial_{\nu}h)(\partial^{\mu}\pi_{i}\partial^{\nu}\pi_{i})
\\
\OHd &=(\partial^{\mu}\pi_{i}\partial_{\mu}\pi_{i})(\partial^{\nu}\pi_{j}\partial_{\nu}\pi_{j})
\\
\OHe &=(\partial^{\mu}\pi_{i}\partial^{\nu}\pi_{i})(\partial_{\mu}\pi_{j}\partial_{\nu}\pi_{j}).
\label{eq:HEFT}
\eea
While power counting is subtle in the HEFT~\cite{Gavela:2016bzc,Alonso:2015fsp}, in our case, where we are interested in the four-derivative operators that appear in subtracted dispersion relations~\cite{Adams:2006sv}, the operators in Eq.~\eqref{eq:HEFT} are precisely those that contribute.
HEFT operators are more commonly defined from a CCWZ construction~\cite{Coleman:1969sm,Callan:1969sn}, $\exp[\pi_i \tau_i/v]$, with $\tau_i$ the three generators of O(4) broken by the Higgs vev.
Our operator basis can be written in an unbroken, O(4) invariant way~\footnote{Let us briefly expand on this point.
We write the Higgs doublet in terms of four real scalar fields as
$$
H = \frac{1}{\sqrt{2}} \begin{pmatrix} \phi_2 + i \phi_1 \\ \phi_4 - i \phi_3 \end{pmatrix}
$$
and associate the vev with $\langle \phi_I \rangle = v \,\delta_{I4}$.
In the notation of Ref.~\cite{Alonso:2016oah}, we can then decompose $H$ into polar coordinates determined by a radius, $v+h$, and unit direction $\bf n$ on the 3-sphere,  constructed such that $\phi_I = (v+h) n_I$.
Being a unit vector, we have ${\bf n} \cdot {\bf n} = 1$, and it is parameterized by the three angular coordinates $\pi_i/v$, in detail 
$$
{\bf n} = \begin{pmatrix} 0 \\ 0 \\ 0 \\ 1 \end{pmatrix} \exp[\pi_i \tau_i/v],
$$
where again $\tau_i$ are the generators of the rotations broken when the Higgs vev reduces custodial symmetry from ${\rm O}(4)\rightarrow {\rm O}(3)$.
Using this language, the operators in the HEFT can be explicitly written in terms of ${\bf n}$ rather than $\pi_i$.
As a single example, we have $\mathcal{O}_1^{h\pi} = v^2 (\partial h)^2(\partial_{\mu}{\bf n}\cdot\partial^{\mu}{\bf n})$, up to terms of higher multiplicity in fields.}, but as positivity bounds are computed from scattering amplitudes, the above basis in terms of asymptotic states is more convenient.
In principle, operators of lower order in field multiplicity or derivatives could also contribute to the amplitudes we study.
At present, since for dispersion relations we are sensitive to the part of the IR amplitude quartic in momenta, we ignore such operators by assuming a weakly coupled completion in which loops or multiple insertions of EFT operators are suppressed, though this would be interesting to generalize.
While it is possible that running of couplings from loops of light states can eventually dominate over tree-level insertions in the deep IR, the assumption of weak coupling by definition implies the existence of some kinematic region---below the UV scale but not in the exponentially far IR~\cite{Arkani-Hamed:2021ajd}---in which the contact insertions of Wilson coefficient computed at tree level will be dominant, which we take to be our regime of interest.

Further, HEFT operators in principle need only be suppressed by the Higgs vev $v$, although given the lack of clear new physics signals, we take our UV scale $\Lambda_{\rm UV} \gg v$, which ensures that our theory satisfies perturbative unitarity~\cite{Cohen:2021ucp} and allows us to treat the Higgs fields as effectively massless throughout.
More generally, in differentiating HEFT from SMEFT, we are making the distinction at fixed order in the expansion in derivatives and multiplicities; of course, at all orders in the expansion, HEFT and SMEFT should coincide.
As a result, certain higher-order SMEFT terms could lead to apparent violation of fixed order SMEFT positivity bounds.
Definitive conclusions about the suitability of HEFT versus SMEFT as the proper low-energy description will of course require multiple measurements to probe the validity of the EFT expansion.
In this work, we restrict ourselves to quartic order in derivatives and fields, to illustrate how, even at fixed order, positivity bounds illustrate a difference between HEFT and SMEFT.
Therefore, at fixed order HEFT$\,\supset\,$SMEFT~\footnote{More explicitly, the Wilson coefficients of SMEFT form a self-contained vector subspace within the space of HEFT coefficients.
While a violation of the SMEFT bounds, and the observation of nonzero Wilson coefficients in the HEFT-allowed region of \Fig{fig:proj}, would arguably correspond to a smoking gun for the HEFT, the converse is not true; it is possible for the correct EFT to be the HEFT, but for its Wilson coefficients to be such that they do not allow the theory to be distinguished from the SMEFT via positivity bounds, e.g., if they land in the SMEFT-allowed region of \Fig{fig:proj}.}, and a particular two-dimensional slice of \Eq{eq:HEFT} reduces to \Eq{eq:SMEFTcustodial} (cf. Ref.~\cite{Salas-Bernardez:2022hqv}).
Since
\bea
\OSa &= \frac{1}{4}(\OHa+2\OHc+\OHe)
\\\OSb &= \frac{1}{4}(\OHa+2\OHb+\OHd),
\label{eq:SMEFTHEFT}
\eea
the slice is defined by 
\be 
4 \cHa\,{=}\,\cSa\,{+}\,\cSb,\;2\cHb\,{=}\,4 \cHd \,{=}\, \cSb,\; 2\cHc \,{=}\, 4 \cHe \,{=}\, \cSa.\label{eq:subspaceconstraints}
\ee

%%%%%%%%%%%%%%%%%%%%%%%%%%%%%%%%%
\medskip
\mysec{Positivity Bounds}
%%%%%%%%%%%%%%%%%%%%%%%%%%%%%%%%%
%
We next determine the bounds on the SMEFT and HEFT operators that result from demanding that the UV be unitary, local, and causal.
For the SMEFT, the required bounds have already been established~\cite{Remmen:2019cyz,ZZ}.
Restricting to the custodial sector in \Eq{eq:SMEFTcustodial}, those results become~\footnote{Throughout, we write all inequalities as strict for brevity.
However, if we are working at fixed order in the coupling expansion in a weakly coupled theory, all of these results reduce to weak inequalities~\cite{Adams:2006sv,Nicolis:2009qm,Chandrasekaran:2018qmx,Remmen:2019cyz}.}
\be
\cSa >0\hspace{0.2cm}
{\rm and}\hspace{0.2cm}
\cSa + \cSb > 0.
\label{eq:SMEFTcustodialbounds} 
\ee
These restrictions are shown in \Fig{fig:proj}.

The equivalent HEFT bounds have not been constructed.
As a first step to doing so, we construct the two-to-two scattering amplitudes mediated by the operators in \Eq{eq:HEFT}.
We consider the most general elastic scattering processes with the incident states constructed from arbitrary superpositions of the $\pi_i$ and $h$.
In detail, we take one initial state to be $\vert 1 \rangle = \alpha_i \vert \pi_i \rangle + \alpha_h \vert h \rangle$, with the four coefficients normalized by ${\boldsymbol \alpha}^2 + \alpha_h^2 = 1$, writing $\boldsymbol \alpha$ for the vector $\alpha_i$.
Without loss of generality, we can choose the overall sign so that $\alpha_h > 0$.
As we have real scalars, the four coefficients can be taken to be real.
The second initial state is defined similarly, but with $({\boldsymbol \beta},\beta_h)$.

It is convenient to arrange the fields into a multiplet $\Phi_I = (\pi_i,h)$, so that $\Phi_4 = h$.
Doing so, the HEFT operators can be combined into $c_{IJKL}\partial_{\mu}\Phi_{I}\partial^{\mu}\Phi_{J}\partial_{\nu}\Phi_{K}\partial^{\nu}\Phi_{L}$, where by definition $c_{IJKL}=c_{JIKL}=c_{IJLK}=c_{KLIJ}$.
In terms of the Wilson coefficients defined in \Eq{eq:HEFT}, 
\bea
c_{IJKL}=& \,\cHa\delta_{I4}\delta_{J4}\delta_{K4}\delta_{L4}+\cHc\delta_{4(I}\bar{\delta}_{J)(K}\delta_{L)4}
\\
&+\frac{1}{2}\cHb(\delta_{I4}\delta_{J4}\bar{\delta}_{KL}+\bar{\delta}_{IJ}\delta_{K4}\delta_{L4})
\\
& +\cHd\bar{\delta}_{IJ}\bar{\delta}_{KL} +\cHe\bar{\delta}_{I(K}\bar{\delta}_{L)J},
\label{eq:Wilson}
\eea
where for brevity we write $\bar{\delta}_{IJ}=\delta_{IJ}-\delta_{I4}\delta_{J4}$.

We next compute the forward elastic scattering amplitude, $A(s)$, which is a function purely of the center-of-mass energy squared (Mandelstam $s$), as in the forward limit we have vanishing exchanged momentum, so that Mandelstam $t \to 0$.
As we review shortly, positivity can be related to the $s^2$ coefficient of the forward amplitude, which in terms of the general notation above is given by $A''(s) = A''_{IJKL}\alpha_{I}\beta_{J}\alpha_{K}\beta_{L}$,
where
\be
A''_{IJKL} = 4(c_{IJKL}+c_{ILKJ}).
\label{eq:ddApre}
\ee
Using \Eq{eq:Wilson} and writing $\alpha = |{\boldsymbol \alpha}|$ and $\beta = |{\boldsymbol \beta}|$,
\bea
A''(s) =&\,
8\cHa (1\,{-}\,\alpha^2)(1\,{-}\,\beta^2)\,{+}\,2\cHc(\alpha^2{+}\,\beta^2{-}\,2\alpha^2\beta^2)\!\!\!\\&+4(2\cHb+\cHc) \sqrt{(1-\alpha^2)(1-\beta^2)} ({\boldsymbol \alpha}\cdot{\boldsymbol \beta})\\&+8(\cHd\,{+}\,\cHe)({\boldsymbol \alpha}\cdot{\boldsymbol \beta})^2 +4\cHe({\boldsymbol \alpha} \times {\boldsymbol \beta})^2.
\label{eq:ddA}
\eea

This result is primed for positivity.
In particular, when analytically continued to complex $s$, $A(s)$ is an analytic function up to discontinuities along the real axis associated with single- or multi-particle exchanges in the $s$- and $u$-channels~\cite{Froissart:1961ux,Martin:1962rt,Martin:1965jj,Mandelstam:1958xc,Lehmann:1958ita}.
One can therefore use contour integration relate the EFT amplitude $A''(s)$ to the UV cross section by connecting a contour at small $|s|$ to one at large $|s|$.
In detail, one finds the classic result~\cite{Adams:2006sv}
\bea
A''(s) = \frac{4}{\pi}\int_0^\infty \frac{{\rm d}s}{s^2} \sigma(s) > 0.
\label{eq:dispersion}
\eea
See Ref.~\cite{Remmen:2019cyz} for further review.
Here we simply emphasize that the result only holds if the UV is unitary and local, using locality to invoke analyticity of the amplitude off of the real axis and thereby deform the contour and unitarity to invoke the optical theorem connecting the discontinuity across the axis to the cross section.

Once positivity in \Eq{eq:dispersion} is established, \Eq{eq:ddA}  implies a set of constraints on the HEFT coefficients.
Positivity must hold for {\it any} ${\boldsymbol \alpha}$ and ${\boldsymbol \beta}$ satisfying $\alpha,\beta \leq 1$.
Marginalizing over all possible choices, we find the following succinct set of positivity bounds,
\bea
\cHa & >0, \qquad \cHc >0,\\
\cHd+\cHe & >0,\qquad \cHe\;\, >0,\\
-\cHc-\sqrt{4\cHa(\cHd+\cHe)}&<\cHb  <\sqrt{4\cHa(\cHd+\cHe)}.
\label{eq:bounds}
\eea

A careful derivation of these bounds is provided in the Appendix.
There we further show that the constraints are actually more general than they naively appear.
In particular, the results in \Eq{eq:bounds} are derived from the optical theorem applied to elastic scattering.
However, for general EFTs it is possible to obtain even stronger bounds using the {\it generalized} optical theorem, as discussed in Refs.~\cite{ZZ,YZZ,Trott,Arkani-Hamed:2021ajd,HYZZ,Freytsis:2022aho}.
For the HEFT, however, we show that the generalized optical theorem yields {\it precisely} the same bounds as in \Eq{eq:bounds}.

%%%%%%%%%%%%%%%%%%%%%%%%%%%%%%%%%
\medskip
\mysec{Positivity and Projection}
%%%%%%%%%%%%%%%%%%%%%%%%%%%%%%%%%
%
We next consider the implications of the newly derived bounds for the SMEFT parameter space in \Fig{fig:proj}.
If we simply take the SMEFT slice of the HEFT using  \Eq{eq:subspaceconstraints}, then \Eq{eq:bounds} collapses exactly to the SMEFT constraints of Eq.~\eqref{eq:SMEFTcustodialbounds}.
But this conclusion would too quick: the shadow cast by the HEFT onto this plane could be larger.
We instead need to project the positive HEFT onto the SMEFT subspace.

To perform this projection, we first observe that the SMEFT subspace is defined by three constraints, $\cHb-2\cHd=0$, $\cHc-2\cHe=0$, and $2\cHa-\cHb-\cHc=0$, which we write in matrix form as $V_{ij} c_j = 0$, where
\be
V_{ij} = \frac{1}{\sqrt{35}}\begin{pmatrix} 0 & -\sqrt{7} & 0 & 2\sqrt{7} & 0 \\  0  & 0 & -\sqrt{7} & 0 & 2\sqrt{7} \\ -5 & 2 & 2 & 1 & 1 \end{pmatrix}\!,
\ee
defining the labels $c_i\,{=}\,(\cHa,\cHb,\cHc,\cHd,\cHe)$.
The constraints are invariant under adding any linear combinations of the rows of $V$, and we have used this freedom to ensure $V_{ij} V_{kj}=\delta_{ik}$. That is, $V_{1j}$, $V_{2j}$, and $V_{3j}$ define a basis of  orthonormal vectors perpendicular to the SMEFT plane.
The projection of HEFT coefficients $c_k$ onto the SMEFT plane is then given by $\hat c_k = c_k - V_{ij}c_j V_{ik}$.
We can thus decompose completely general HEFT coefficients $c_k$ into a contribution within and perpendicular to the SMEFT plane, as
\be 
c_k=\hat c_k + d_i V_{ik}
\label{eq:ddef}
\ee
for some $d_{1,2,3}$.
Using \Eq{eq:subspaceconstraints}, we write the $\hat c_k$ in terms of the SMEFT coefficients $C_{+,\times}$,
\be
\hat c_k = \frac{1}{4}\left(\cSa\,{+}\,\cSb,\;2\cSb,\;2\cSa,\;\cSb,\;\cSa\right).
\label{eq:ckdef}
\ee

On the slice of the HEFT that corresponds to the SMEFT, we have $d_{1,2,3} = 0$, and as noted, the general bounds reduce to those of the SMEFT.
In other words, if low-energy physics is described by the SMEFT, and we parameterize physics in terms of the HEFT and then postselect down to the SMEFT, the constraints are the same as if we had used the SMEFT all along, as expected.

\begin{figure}[t]
\begin{center}
\includegraphics[width=7.6cm]{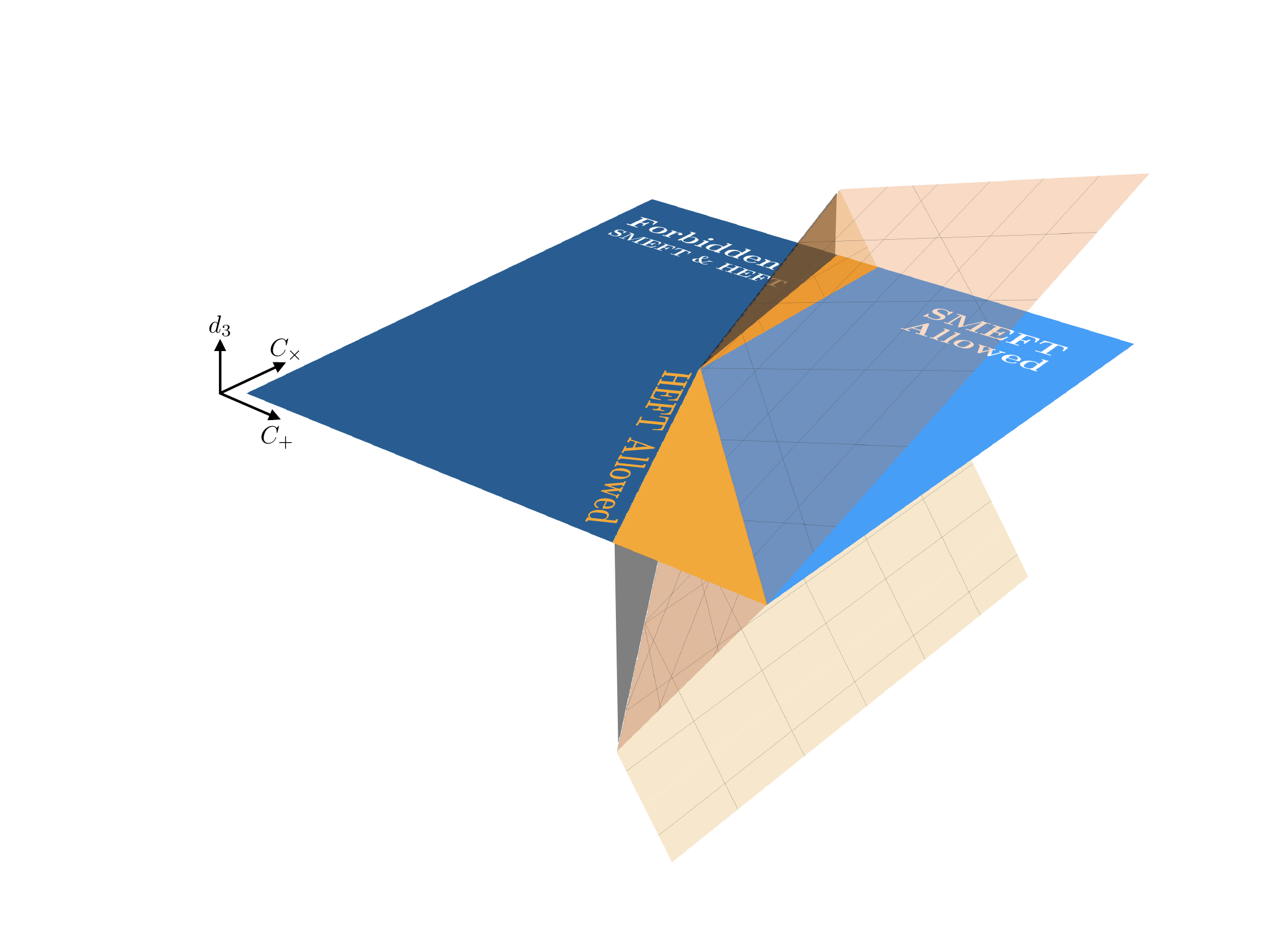}
\end{center}
\vspace{-3mm}
\caption{A perspective on the projection of the HEFT to the SMEFT.
Beyond the SMEFT plane shown in \Fig{fig:proj}, we have included the additional dimension $d_3$ of the HEFT; we continue to marginalize over $d_{1,2}$.
The space permitted by HEFT positivity bounds---that is, where there exist some $d_{1,2}$ such that a given point $(\cSa,\cSb,d_3)$ complies with the HEFT positivity bounds---is to the right of the hatched light orange contour.
The contour provides an indication of how the HEFT can enlarge the allowed parameters to the orange region, and in dark gray we illustrate the projection to the SMEFT. \vspace{-3mm}
}
\label{fig:proj3D}
\end{figure}

What about the converse?
That is, what if the low-energy physics of our universe is described by the HEFT, but we instead naively use the SMEFT in defining our positivity bounds and experimentally measuring deviations from the SM?
In that case, the allowed space of bounds in the SMEFT plane is that for which there {\it exist} some $d_{1,2,3}$  for which the point in the SMEFT plane satisfies the HEFT bounds in \Eq{eq:bounds}.
In other words, using the SMEFT in a world defined by the HEFT, the space of $C_{+,\times}$ consistent with unitarity and locality is the {\it projection} of the five-dimensional HEFT cone onto the SMEFT plane, rather than the slice of the cone through the plane.
Inputting the parameterization of the $c_k$ in \Eq{eq:ddef} into the HEFT bounds and marginalizing over the $d_{1,2,3}$, we find that the projected bounds become
\be
6 \cSa + \cSb >0\hspace{0.2cm}
{\rm and}\hspace{0.2cm}
19 \cSa +9 \cSb >0.
\label{eq:boundsproj}
\ee
The region allowed by the bounds for the SMEFT itself in \Eq{eq:SMEFTcustodialbounds} forms a {\it strict subset} of the region permitted by \Eq{eq:boundsproj}. 
If the universe is described by the HEFT and not the SMEFT, but one erroneously parameterizes new physics in terms of the SMEFT anyway, then one could see an apparent violation of positivity in the SMEFT coefficients simply due to this projection.
This justifies the HEFT region depicted in \Fig{fig:proj}; see \Fig{fig:proj3D} for a perspective on how the additional parameter space emerges.

%%%%%%%%%%%%%%%%%%%%%%%%%%%%%%%%%
\medskip
\mysec{Ultraviolet Extensions}
%%%%%%%%%%%%%%%%%%%%%%%%%%%%%%%%%
%
As an illustration that our bounds capture realistic scenarios for new physics, let us write down UV extensions of the HEFT under which the bounds in \Eq{eq:bounds} are saturated.
That is, we consider tree-level models involving a single massive state coupled to a bilinear of the light Higgs fields, where integrating out the new heavy field gives the operators in \Eq{eq:HEFT}.
The phrasing ``UV extensions'' indicates that the theories need not by explicitly UV complete so long as they parametrically raise the cutoff of the theory.
For example, if we have a scalar $X$ of mass $m$ that couples to $h$ via the dimension-five operator $(X/\Lambda_X)(\partial h)^2$, for some scale $\Lambda_X \gg m$, then integrating out $X$ generates the dimension-eight HEFT operator $(\partial h)^4$ with Wilson coefficient $\cHa = 1/2\Lambda_X^2 m^2$.
That is, integrating in $X$ raises the HEFT cutoff from $\Lambda_{\rm UV} = \sqrt{2\Lambda_X m}$ up to $\Lambda_X\gg \Lambda_{\rm UV}$, so this model is a UV extension of the HEFT~\footnote{See Refs.~\cite{Remmen:2019cyz,ZZ} for discussion of tree-level UV extensions of the SMEFT Higgs operators.}.

Let us consider a theory containing the following massive particles: scalars $X$ and $Y$ transforming as singlets under ${\rm O}(3)$, a scalar $Z_i$ transforming as a $\bf 3$ (vector) of ${\rm O}(3)$, and two massive spacetime vectors $A_i^\mu$ and $B_i^\mu$ also transforming as $\bf 3$ (vectors) of ${\rm O}(3)$.
We couple these states to the Higgs in a custodial invariant Lagrangian,
\bea
\hspace{-1mm}
&{\cal L}\supset g_{1}X\partial_{\mu}h\partial^{\mu}h+g_{2}X\partial_{\mu}\pi_i\partial^{\mu}\pi_i+g_{3}Y\partial_{\mu}h\partial^{\mu}h 
\\
&{+}g_{4} Z_i\partial_{\mu}\pi_i \partial^{\mu}h{+} m g_{5}h A_i^{\mu}\partial_{\mu}\pi_i {+}mg_{6}\epsilon_{ijk}B^{\mu}_{i}\pi_{j}\partial_{\mu}\pi_k,
\label{eq:UV}
\eea
taking all heavy particles to have mass $m$. We have introduced factors of $m$ such that all of the couplings $g_i$ have mass dimension $-1$.
Just as for the HEFT basis, the interactions in Eq.~\eqref{eq:UV} can be straightforwardly written with explicit O(4) invariance~\footnote{That is, all of these interactions can be simply written as couplings of heavy states to the unit vector ${\bf n}$.
The only subtlety in writing the ${\rm O}(4)$ analogue of our UV extensions is that for the $B$ vector one first needs to dualize to an ${\rm O}(4)$ valued two-form, i.e., $B^{ab}_{\mu}n^a\partial^\mu n^b$.}.
Integrating out the UV states, we have the following Wilson coefficients for the HEFT operators in~\Eq{eq:HEFT},
\bea
{c}_{1} & =\hat g_{1}^{2}+ \hat g_{3}^{2},\;\; & {c}_{2} & =2\hat g_{1}\hat g_{2}-\hat g_{5}^{2},\;\; &
{c}_{3} & =\hat g_{4}^{2}+\hat g_{5}^{2}\\
{c}_{4} & =\hat g_{2}^{2}-2\hat g_{6}^{2},&
{c}_{5} & =2\hat g_{6}^{2},
\eea
for brevity defining $\hat g_i = g_i/\sqrt{2}m$, in addition to the dimension-six terms,
\be
{\cal L}\supset\frac{1}{2}g_{5}^{2}h^{2}(\partial_\mu \pi_i \partial^\mu \pi_i)+g_{6}^{2}g_{\mu\nu}(\pi_{[i} \partial^\mu \pi_{j]})(\pi_{[i}\partial^\nu \pi_{j]}).
\ee
The region spanned by the couplings $\hat g_{i}$ within the space of Wilson coefficients is precisely that given by the positivity bounds in \Eq{eq:bounds}.
That is, we have found tree-level UV extensions of the HEFT that span the full space of coefficients consistent with unitarity.

A simple UV extension of the HEFT that violates SMEFT positivity is the interaction $g X\partial_\mu \pi_i \partial^\mu \pi_i$, that is, \Eq{eq:UV} in the case where the couplings satisfy $g_i = g \times (0,1,0,0,0,0)$.
This UV extension of the HEFT via a single scalar field, reminiscent of a sigma model, generates Wilson coefficients that, when projected down to the SMEFT plane, live on the line $6\cSa + \cSb = 0$, with $\cSa < 0$, violating SMEFT positivity~\eqref{eq:SMEFTcustodialbounds} (but satisfying $19 \cSa + 9 \cSb > 0$ from HEFT~\eqref{eq:boundsproj}).
Similarly, the interaction $g X(\partial_\mu h \partial^\mu h - 2\partial_\mu \pi_i \partial^\mu \pi_i)$, i.e., $g_i = g \times (1,-2,0,0,0,0)$, also reminiscent of a sigma model, generates Wilson coefficients after projection living on the line $19\cSa + 9\cSb = 0$, but with $\cSa > 0$, thereby violating the SMEFT bounds~\eqref{eq:SMEFTcustodialbounds} with $\cSa + \cSb < 0$ (but satisfying $6\cSa + \cSb > 0$ from HEFT~\eqref{eq:boundsproj}). 
That is, these two simple scalar extensions of the HEFT generate the two rays in \Fig{fig:proj} separating the orange HEFT-only region from the dark blue forbidden region.
We reiterate that the scenarios studied here are simple UV extensions; we leave open the important question of how generic it is for UV complete models to populate the HEFT window.

%%%%%%%%%%%%%%%%%%%%%%%%%%%%%%%%%
\medskip
\mysec{Discussion}
%%%%%%%%%%%%%%%%%%%%%%%%%%%%%%%%%
%
The discovery of a nonvanishing SMEFT Wilson coefficient would herald the first clear sign at a collider of the breakdown of the SM.
The promise of positivity is that it can enhance any such discovery into a probe of the principles governing the emerging UV: is the theory unitary, local, and causal?
However, as we have shown in this Letter, these searches also probe the symmetry structure of the UV, and a discovery in the orange region in \Fig{fig:proj} would be a strong indication that the new physics is best described by the HEFT.

By assuming custodial symmetry and performing a purely tree-level analysis, the present work represents a first glimpse into positivity applied to the HEFT.
Moving forward, it would be important to understand the impact of multiple insertions of operators of lower mass dimension that arise when considering loops; there is a growing appreciation that such contributions can play a key role in dispersive analyses~\cite{Bellazzini:2020cot,Chala:2021wpj,Chala:2021pll,Li:2022aby,Liao:2025npz}.
Further, it will be important to identify and analyze complete UV models that can populate the HEFT allowed region in \Fig{fig:proj} to provide context for future experimental searches.
Such models will provide a testing ground for understanding the importance of various simplifying assumptions adopted in our work (such as neglecting loop effects) and on to what extent a discovery is within reach when constraints from other searches are incorporated (see e.g. Ref.~\cite{Cappati:2022skp}).

One of the challenges in studying EFTs is identifying interesting regions of the vast parameter space; the SMEFT alone has 44,807 operators at dimension eight~\cite{Henning:2015alf}.
Although simplifying assumptions have been adopted, this Letter demonstrates that the two-dimensional space of \Fig{fig:proj} appears rich.
It is also a space well poised to be further tested at the High Luminosity run at the Large Hadron Collider.
Both ATLAS~\cite{ATLAS:2018mxa,ATLAS:2019thr,ATLAS:2019qhm,ATLAS:2020nlt,ATLAS:2023sua,ATLAS:2024ini} and CMS~\cite{CMS:2019uys,CMS:2019mpq,CMS:2019qfk,CMS:2020ioi,CMS:2020gfh,CMS:2020fqz,CMS:2020ypo,CMS:2021jji,CMS:2021gme,CMS:2022dmc,CMS:2022yrl} have performed analyses searching for aQGCs induced by the SMEFT at dimension eight, including explicit studies of two-dimensional slices of this parameter space~\cite{CMS:2019uys,ATLAS:2023sua,ATLAS:2024ini}.
The space of \Fig{fig:proj} is therefore ready to be explored and may yet reveal the first hints of SMEFT or even HEFT.

%%%%%%%%%%%%%%%%%%%%%%%%%%%%%%%%%
\medskip
\noindent {\it Acknowledgments.}
%%%%%%%%%%%%%%%%%%%%%%%%%%%%%%%%%
%
We greatly appreciate comments provided by Tim Cohen on a draft version of this work.
G.N.R. is supported by the James Arthur Postdoctoral Fellowship at New York University.
The work of N.L.R. was supported by the Office of High Energy Physics of the U.S. Department of Energy under contract DE-AC02-05CH11231.

\vspace{-5mm}
\bibliographystyle{utphys-modified}
\bibliography{HEFTpositivity}

\providecommand{\href}[2]{#2}\begingroup\raggedright\begin{thebibliography}{100}

\bibitem{ATLAS:2023sua}
{\bfseries ATLAS} {\bfseries Collaboration}, G.~Aad { et~al.}, ``{Measurement
  and interpretation of same-sign W boson pair production in association with
  two jets in pp collisions at $ \sqrt{s} $ = 13 TeV with the ATLAS
  detector},'' \href{http://dx.doi.org/10.1007/JHEP04(2024)026}{{\em JHEP}
  {\bfseries 04} (2024) 026}, \href{http://arxiv.org/abs/2312.00420}{{\ttfamily
  arXiv:2312.00420 [hep-ex]}}.

\bibitem{Adams:2006sv}
A.~Adams, N.~Arkani-Hamed, S.~Dubovsky, A.~Nicolis, and R.~Rattazzi,
  ``{Causality, analyticity and an IR obstruction to UV completion},''
  \href{http://dx.doi.org/10.1088/1126-6708/2006/10/014}{{\em JHEP} {\bfseries
  10} (2006) 014}, \href{http://arxiv.org/abs/hep-th/0602178}{{\ttfamily
  arXiv:hep-th/0602178}}.

\bibitem{Jenkins:2006ia}
A.~Jenkins and D.~O'Connell, ``{The Story of ${\cal O}$: Positivity constraints
  in effective field theories},''
  \href{http://arxiv.org/abs/hep-th/0609159}{{\ttfamily arXiv:hep-th/0609159}}.

\bibitem{Dvali:2012zc}
G.~Dvali, A.~Franca, and C.~Gomez, ``{Road Signs for UV-Completion},''
  \href{http://arxiv.org/abs/1204.6388}{{\ttfamily arXiv:1204.6388 [hep-th]}}.

\bibitem{Nicolis:2009qm}
A.~Nicolis, R.~Rattazzi, and E.~Trincherini, ``{Energy's and amplitudes'
  positivity},'' \href{http://dx.doi.org/10.1007/JHEP05(2010)095}{{\em JHEP}
  {\bfseries 05} (2010) 095}, \href{http://arxiv.org/abs/0912.4258}{{\ttfamily
  arXiv:0912.4258 [hep-th]}}.
[Erratum: \href{https://doi.org/10.1007/JHEP11(2011)128}{{\it JHEP} {\bf 11}
  (2011) 128}].
%%CITATION = ARXIV:0912.4258;%%.

\bibitem{Arkani-Hamed:2020blm}
N.~Arkani-Hamed, T.-C. Huang, and Y.-t. Huang, ``{The EFT-Hedron},''
  \href{http://dx.doi.org/10.1007/JHEP05(2021)259}{{\em JHEP} {\bfseries 05}
  (2021) 259}, \href{http://arxiv.org/abs/2012.15849}{{\ttfamily
  arXiv:2012.15849 [hep-th]}}.

\bibitem{Bellazzini:2020cot}
B.~Bellazzini, J.~Elias~Mir\'o, R.~Rattazzi, M.~Riembau, and F.~Riva,
  ``{Positive moments for scattering amplitudes},''
  \href{http://dx.doi.org/10.1103/PhysRevD.104.036006}{{\em Phys. Rev. D}
  {\bfseries 104} (2021) 036006},
  \href{http://arxiv.org/abs/2011.00037}{{\ttfamily arXiv:2011.00037
  [hep-th]}}.

\bibitem{Note1}
For instance, among many possible examples ranging from phenomenology to formal
  theory, positivity has been applied to quantum gravitational modifications of
  general relativity and their implications for the Weak Gravity
  Conjecture~\cite
  {Bellazzini:2015cra,Cheung:2016wjt,Camanho:2014apa,Gruzinov:2006ie,Arkani-Hamed:2021ajd,Cheung:2014ega,Bellazzini:2019xts,Andriolo:2020lul,Caron-Huot:2021rmr,Caron-Huot:2022ugt,Caron-Huot:2022jli,Cheung:2016yqr,deRham:2017xox,Camanho:2016opx}.

\bibitem{Remmen:2019cyz}
G.~N. Remmen and N.~L. Rodd, ``{Consistency of the Standard Model Effective
  Field Theory},'' \href{http://dx.doi.org/10.1007/JHEP12(2019)032}{{\em JHEP}
  {\bfseries 12} (2019) 032}, \href{http://arxiv.org/abs/1908.09845}{{\ttfamily
  arXiv:1908.09845 [hep-ph]}}.

\bibitem{Remmen:2020vts}
G.~N. Remmen and N.~L. Rodd, ``{Flavor Constraints from Unitarity and
  Analyticity},'' \href{http://dx.doi.org/10.1103/PhysRevLett.127.149901}{{\em
  Phys. Rev. Lett.} {\bfseries 125} (2020) 081601},
  \href{http://arxiv.org/abs/2004.02885}{{\ttfamily arXiv:2004.02885
  [hep-ph]}}. [Erratum:
  \href{https://doi.org/10.1103/PhysRevLett.127.149901}{{\it Phys. Rev. Lett.}
  {\bf 127}, 149901 (2021)}].

\bibitem{Remmen:2022orj}
G.~N. Remmen and N.~L. Rodd, ``{Spinning sum rules for the dimension-six
  SMEFT},'' \href{http://dx.doi.org/10.1007/JHEP09(2022)030}{{\em JHEP}
  {\bfseries 09} (2022) 030}, \href{http://arxiv.org/abs/2206.13524}{{\ttfamily
  arXiv:2206.13524 [hep-ph]}}.

\bibitem{Bi:2019phv}
Q.~Bi, C.~Zhang, and S.-Y. Zhou, ``{Positivity constraints on aQGC: carving out
  the physical parameter space},''
  \href{http://dx.doi.org/10.1007/JHEP06(2019)137}{{\em JHEP} {\bfseries 06}
  (2019) 137}, \href{http://arxiv.org/abs/1902.08977}{{\ttfamily
  arXiv:1902.08977 [hep-ph]}}.

\bibitem{Zhang:2018shp}
C.~Zhang and S.-Y. Zhou, ``{Positivity bounds on vector boson scattering at the
  LHC},'' \href{http://dx.doi.org/10.1103/PhysRevD.100.095003}{{\em Phys. Rev.
  D} {\bfseries 100} (2019) 095003},
  \href{http://arxiv.org/abs/1808.00010}{{\ttfamily arXiv:1808.00010
  [hep-ph]}}.

\bibitem{Remmen:2020uze}
G.~N. Remmen and N.~L. Rodd, ``{Signs, spin, SMEFT: Sum rules at dimension
  six},'' \href{http://dx.doi.org/10.1103/PhysRevD.105.036006}{{\em Phys. Rev.
  D} {\bfseries 105} (2022) 036006},
  \href{http://arxiv.org/abs/2010.04723}{{\ttfamily arXiv:2010.04723
  [hep-ph]}}.

\bibitem{Low:2009di}
I.~Low, R.~Rattazzi, and A.~Vichi, ``{Theoretical Constraints on the Higgs
  Effective Couplings},'' \href{http://dx.doi.org/10.1007/JHEP04(2010)126}{{\em
  JHEP} {\bfseries 04} (2010) 126},
  \href{http://arxiv.org/abs/0907.5413}{{\ttfamily arXiv:0907.5413 [hep-ph]}}.

\bibitem{Englert:2019zmt}
C.~Englert, G.~F. Giudice, A.~Greljo, and M.~Mccullough, ``{The
  $\hat{H}$-Parameter: An Oblique Higgs View},''
  \href{http://dx.doi.org/10.1007/JHEP09(2019)041}{{\em JHEP} {\bfseries 09}
  (2019) 041}, \href{http://arxiv.org/abs/1903.07725}{{\ttfamily
  arXiv:1903.07725 [hep-ph]}}.

\bibitem{ZZ}
C.~Zhang and S.-Y. Zhou, ``{Convex Geometry Perspective on the (Standard Model)
  Effective Field Theory Space},''
  \href{http://dx.doi.org/10.1103/PhysRevLett.125.201601}{{\em Phys. Rev.
  Lett.} {\bfseries 125} (2020) 201601},
  \href{http://arxiv.org/abs/2005.03047}{{\ttfamily arXiv:2005.03047
  [hep-ph]}}.

\bibitem{YZZ}
K.~Yamashita, C.~Zhang, and S.-Y. Zhou, ``{Elastic positivity vs extremal
  positivity bounds in SMEFT: a case study in transversal electroweak
  gauge-boson scatterings},''
  \href{http://dx.doi.org/10.1007/JHEP01(2021)095}{{\em JHEP} {\bfseries 01}
  (2021) 095}, \href{http://arxiv.org/abs/2009.04490}{{\ttfamily
  arXiv:2009.04490 [hep-ph]}}.

\bibitem{Trott}
T.~Trott, ``{Causality, unitarity and symmetry in effective field theory},''
  \href{http://dx.doi.org/10.1007/JHEP07(2021)143}{{\em JHEP} {\bfseries 07}
  (2021) 143}, \href{http://arxiv.org/abs/2011.10058}{{\ttfamily
  arXiv:2011.10058 [hep-ph]}}.

\bibitem{HYZZ}
X.~Li, H.~Xu, C.~Yang, C.~Zhang, and S.-Y. Zhou, ``{Positivity in Multifield
  Effective Field Theories},''
  \href{http://dx.doi.org/10.1103/PhysRevLett.127.121601}{{\em Phys. Rev.
  Lett.} {\bfseries 127} (2021) 121601},
  \href{http://arxiv.org/abs/2101.01191}{{\ttfamily arXiv:2101.01191
  [hep-ph]}}.

\bibitem{Chala:2023xjy}
M.~Chala and X.~Li, ``{Positivity restrictions on the mixing of dimension-eight
  SMEFT operators},'' \href{http://dx.doi.org/10.1103/PhysRevD.109.065015}{{\em
  Phys. Rev. D} {\bfseries 109} (2024) 065015},
  \href{http://arxiv.org/abs/2309.16611}{{\ttfamily arXiv:2309.16611
  [hep-ph]}}.

\bibitem{Hong:2024fbl}
D.-Y. Hong, Z.-H. Wang, and S.-Y. Zhou, ``{On Capped Higgs Positivity Cone},''
\newblock 4, 2024.
\newblock \href{http://arxiv.org/abs/2404.04479}{{\ttfamily arXiv:2404.04479
  [hep-ph]}}.

\bibitem{Chen:2023bhu}
Q.~Chen, K.~Mimasu, T.~A. Wu, G.-D. Zhang, and S.-Y. Zhou, ``{Capping the
  positivity cone: dimension-8 Higgs operators in the SMEFT},''
  \href{http://dx.doi.org/10.1007/JHEP03(2024)180}{{\em JHEP} {\bfseries 03}
  (2024) 180}, \href{http://arxiv.org/abs/2309.15922}{{\ttfamily
  arXiv:2309.15922 [hep-ph]}}.

\bibitem{Feruglio:1992wf}
F.~Feruglio, ``{The Chiral approach to the electroweak interactions},''
  \href{http://dx.doi.org/10.1142/S0217751X93001946}{{\em Int. J. Mod. Phys. A}
  {\bfseries 8} (1993) 4937},
  \href{http://arxiv.org/abs/hep-ph/9301281}{{\ttfamily arXiv:hep-ph/9301281}}.

\bibitem{Bagger:1993zf}
J.~Bagger, V.~D. Barger, K.-m. Cheung, J.~F. Gunion, T.~Han, G.~A. Ladinsky,
  R.~Rosenfeld, and C.~P. Yuan, ``{The Strongly interacting W W system: Gold
  plated modes},'' \href{http://dx.doi.org/10.1103/PhysRevD.49.1246}{{\em Phys.
  Rev. D} {\bfseries 49} (1994) 1246},
  \href{http://arxiv.org/abs/hep-ph/9306256}{{\ttfamily arXiv:hep-ph/9306256}}.

\bibitem{Koulovassilopoulos:1993pw}
V.~Koulovassilopoulos and R.~S. Chivukula, ``{Phenomenology of a nonstandard
  Higgs boson in $W_L W_L$ scattering},''
  \href{http://dx.doi.org/10.1103/PhysRevD.50.3218}{{\em Phys. Rev. D}
  {\bfseries 50} (1994) 3218},
  \href{http://arxiv.org/abs/hep-ph/9312317}{{\ttfamily arXiv:hep-ph/9312317}}.

\bibitem{Alonso:2012px}
R.~Alonso, M.~B. Gavela, L.~Merlo, S.~Rigolin, and J.~Yepes, ``{The Effective
  Chiral Lagrangian for a Light Dynamical `Higgs Particle'},''
  \href{http://dx.doi.org/10.1016/j.physletb.2013.04.037}{{\em Phys. Lett. B}
  {\bfseries 722} (2013) 330}, \href{http://arxiv.org/abs/1212.3305}{{\ttfamily
  arXiv:1212.3305 [hep-ph]}}. [Erratum:
  \href{https://doi.org/10.1016/j.physletb.2013.09.028}{{\it Phys. Lett. B}
  {\bf 726}, 926 (2013)}].

\bibitem{Brivio:2013pma}
I.~Brivio, T.~Corbett, O.~J.~P. \'Eboli, M.~B. Gavela, J.~Gonzalez-Fraile,
  M.~C. Gonzalez-Garcia, L.~Merlo, and S.~Rigolin, ``{Disentangling a dynamical
  Higgs},'' \href{http://dx.doi.org/10.1007/JHEP03(2014)024}{{\em JHEP}
  {\bfseries 03} (2014) 024}, \href{http://arxiv.org/abs/1311.1823}{{\ttfamily
  arXiv:1311.1823 [hep-ph]}}.

\bibitem{Graf:2022rco}
L.~Gr\'af, B.~Henning, X.~Lu, T.~Melia, and H.~Murayama, ``{Hilbert series, the
  Higgs mechanism, and HEFT},''
  \href{http://dx.doi.org/10.1007/JHEP02(2023)064}{{\em JHEP} {\bfseries 02}
  (2023) 064}, \href{http://arxiv.org/abs/2211.06275}{{\ttfamily
  arXiv:2211.06275 [hep-ph]}}.

\bibitem{Cohen:2021ucp}
T.~Cohen, N.~Craig, X.~Lu, and D.~Sutherland, ``{Unitarity violation and the
  geometry of Higgs EFTs},''
  \href{http://dx.doi.org/10.1007/JHEP12(2021)003}{{\em JHEP} {\bfseries 12}
  (2021) 003}, \href{http://arxiv.org/abs/2108.03240}{{\ttfamily
  arXiv:2108.03240 [hep-ph]}}.

\bibitem{Alonso:2015fsp}
R.~Alonso, E.~E. Jenkins, and A.~V. Manohar, ``{A Geometric Formulation of
  Higgs Effective Field Theory: Measuring the Curvature of Scalar Field
  Space},'' \href{http://dx.doi.org/10.1016/j.physletb.2016.01.041}{{\em Phys.
  Lett. B} {\bfseries 754} (2016) 335},
  \href{http://arxiv.org/abs/1511.00724}{{\ttfamily arXiv:1511.00724
  [hep-ph]}}.

\bibitem{Alonso:2016oah}
R.~Alonso, E.~E. Jenkins, and A.~V. Manohar, ``{Geometry of the Scalar
  Sector},'' \href{http://dx.doi.org/10.1007/JHEP08(2016)101}{{\em JHEP}
  {\bfseries 08} (2016) 101}, \href{http://arxiv.org/abs/1605.03602}{{\ttfamily
  arXiv:1605.03602 [hep-ph]}}.

\bibitem{Falkowski:2019tft}
A.~Falkowski and R.~Rattazzi, ``{Which EFT},''
  \href{http://dx.doi.org/10.1007/JHEP10(2019)255}{{\em JHEP} {\bfseries 10}
  (2019) 255}, \href{http://arxiv.org/abs/1902.05936}{{\ttfamily
  arXiv:1902.05936 [hep-ph]}}.

\bibitem{Cohen:2020xca}
T.~Cohen, N.~Craig, X.~Lu, and D.~Sutherland, ``{Is SMEFT Enough?},''
  \href{http://dx.doi.org/10.1007/JHEP03(2021)237}{{\em JHEP} {\bfseries 03}
  (2021) 237}, \href{http://arxiv.org/abs/2008.08597}{{\ttfamily
  arXiv:2008.08597 [hep-ph]}}.

\bibitem{Banta:2021dek}
I.~Banta, T.~Cohen, N.~Craig, X.~Lu, and D.~Sutherland, ``{Non-decoupling new
  particles},'' \href{http://dx.doi.org/10.1007/JHEP02(2022)029}{{\em JHEP}
  {\bfseries 02} (2022) 029}, \href{http://arxiv.org/abs/2110.02967}{{\ttfamily
  arXiv:2110.02967 [hep-ph]}}.

\bibitem{Gomez-Ambrosio:2022why}
R.~G\'omez-Ambrosio, F.~J. Llanes-Estrada, A.~Salas-Bern\'ardez, and J.~J.
  Sanz-Cillero, ``{SMEFT is falsifiable through multi-Higgs measurements (even
  in the absence of new light particles)},''
  \href{http://dx.doi.org/10.1088/1572-9494/ace95e}{{\em Commun. Theor. Phys.}
  {\bfseries 75} no.~9, (2023) 095202},
  \href{http://arxiv.org/abs/2207.09848}{{\ttfamily arXiv:2207.09848
  [hep-ph]}}.

\bibitem{Brivio:2016fzo}
I.~Brivio, J.~Gonzalez-Fraile, M.~C. Gonzalez-Garcia, and L.~Merlo, ``{The
  complete HEFT Lagrangian after the LHC Run I},''
  \href{http://dx.doi.org/10.1140/epjc/s10052-016-4211-9}{{\em Eur. Phys. J. C}
  {\bfseries 76} (2016) 416}, \href{http://arxiv.org/abs/1604.06801}{{\ttfamily
  arXiv:1604.06801 [hep-ph]}}.

\bibitem{Brivio:2017vri}
I.~Brivio and M.~Trott, ``{The Standard Model as an Effective Field Theory},''
  \href{http://dx.doi.org/10.1016/j.physrep.2018.11.002}{{\em Phys. Rept.}
  {\bfseries 793} (2019) 1}, \href{http://arxiv.org/abs/1706.08945}{{\ttfamily
  arXiv:1706.08945 [hep-ph]}}.

\bibitem{Alonso:2023upf}
R.~Alonso, ``{A primer on Higgs Effective Field Theory with Geometry},''
  \href{http://arxiv.org/abs/2307.14301}{{\ttfamily arXiv:2307.14301
  [hep-ph]}}.

\bibitem{Arkani-Hamed:2021ajd}
N.~Arkani-Hamed, Y.-t. Huang, J.-Y. Liu, and G.~N. Remmen, ``{Causality,
  unitarity, and the weak gravity conjecture},''
  \href{http://dx.doi.org/10.1007/JHEP03(2022)083}{{\em JHEP} {\bfseries 03}
  (2022) 083}, \href{http://arxiv.org/abs/2109.13937}{{\ttfamily
  arXiv:2109.13937 [hep-th]}}.

\bibitem{Freytsis:2022aho}
M.~Freytsis, S.~Kumar, G.~N. Remmen, and N.~L. Rodd, ``{Multifield positivity
  bounds for inflation},''
  \href{http://dx.doi.org/10.1007/JHEP09(2023)041}{{\em JHEP} {\bfseries 09}
  (2023) 041}, \href{http://arxiv.org/abs/2210.10791}{{\ttfamily
  arXiv:2210.10791 [hep-th]}}.

\bibitem{Eboli:2006wa}
O.~J.~P. \'Eboli, M.~C. Gonzalez-Garcia, and J.~K. Mizukoshi, ``{$p p
  \rightarrow jj e^\pm \mu^\pm \nu\nu$ and $jje^\pm \mu^\mp \nu\nu$ at ${\cal
  O}(\alpha_{\rm em}^6)$ and ${\cal O}(\alpha_{\rm em}^4 \alpha_s^2)$ for the
  study of the quartic electroweak gauge boson vertex at CERN LHC},''
  \href{http://dx.doi.org/10.1103/PhysRevD.74.073005}{{\em Phys. Rev. D}
  {\bfseries 74} (2006) 073005},
  \href{http://arxiv.org/abs/hep-ph/0606118}{{\ttfamily arXiv:hep-ph/0606118}}.

\bibitem{Durieux:2024zrg}
G.~Durieux, G.~N. Remmen, N.~L. Rodd, O.~J.~P. \'Eboli, M.~C. Gonzalez-Garcia,
  D.~Kondo, H.~Murayama, and R.~Okabe, ``{LHC EFT WG Note: Basis for Anomalous
  Quartic Gauge Couplings},'' \href{http://arxiv.org/abs/2411.02483}{{\ttfamily
  arXiv:2411.02483 [hep-ph]}}.

\bibitem{Note2}
We note that it is never possible for any combination of SMEFT-like and
  HEFT-like UV physics to produce Wilson coefficients that, upon projection to
  the SMEFT, land in the forbidden dark blue region in Fig.~\ref {fig:proj}.
  This is clear since every UV completion that transforms linearly under the SM
  gauge group (producing the SMEFT) can be written as a special case of a HEFT
  completion.

\bibitem{Hays:2018zze}
C.~Hays, A.~Martin, V.~Sanz, and J.~Setford, ``{On the impact of
  dimension-eight SMEFT operators on Higgs measurements},''
  \href{http://dx.doi.org/10.1007/JHEP02(2019)123}{{\em JHEP} {\bfseries 02}
  (2019) 123}, \href{http://arxiv.org/abs/1808.00442}{{\ttfamily
  arXiv:1808.00442 [hep-ph]}}.

\bibitem{Note3}
For a collection of $n$ scalars $\Phi _I$, there are $n(n\protect \,{+}\protect
  \,1)(n^2\protect \,{+}\protect \,n\protect \,{+}\protect \,2)/8$ EFT
  operators of the form $c_{IJKL}\partial _\mu \Phi _I \partial ^\mu \Phi _J
  \partial _\nu \Phi _K \partial ^\nu \Phi _L$. The CP-violating operator
  $\epsilon ^{\mu \nu \rho \sigma } \partial _{\mu } \Phi _{I} \partial _{\nu }
  \Phi _J \partial _{\rho } \Phi _K \partial _{\sigma } \Phi _{K}$ can be
  written as a total derivative upon integration by parts.

\bibitem{Gavela:2016bzc}
B.~M. Gavela, E.~E. Jenkins, A.~V. Manohar, and L.~Merlo, ``{Analysis of
  General Power Counting Rules in Effective Field Theory},''
  \href{http://dx.doi.org/10.1140/epjc/s10052-016-4332-1}{{\em Eur. Phys. J. C}
  {\bfseries 76} (2016) 485}, \href{http://arxiv.org/abs/1601.07551}{{\ttfamily
  arXiv:1601.07551 [hep-ph]}}.

\bibitem{Coleman:1969sm}
S.~R. Coleman, J.~Wess, and B.~Zumino, ``{Structure of phenomenological
  Lagrangians. 1.},'' \href{http://dx.doi.org/10.1103/PhysRev.177.2239}{{\em
  Phys. Rev.} {\bfseries 177} (1969) 2239}.

\bibitem{Callan:1969sn}
C.~G. Callan, Jr., S.~R. Coleman, J.~Wess, and B.~Zumino, ``{Structure of
  phenomenological Lagrangians. 2.},''
  \href{http://dx.doi.org/10.1103/PhysRev.177.2247}{{\em Phys. Rev.} {\bfseries
  177} (1969) 2247}.

\bibitem{Note4}
Let us briefly expand on this point. We write the Higgs doublet in terms of
  four real scalar fields as $$ H = \protect \frac {1}{\protect \sqrt {2}}
  \begin {pmatrix} \phi _2 + i \phi _1 \\ \phi _4 - i \phi _3 \end {pmatrix} $$
  and associate the vev with $\langle \phi _I \rangle = v \protect \,\delta
  _{I4}$. In the notation of Ref.~\cite {Alonso:2016oah}, we can then decompose
  $H$ into polar coordinates determined by a radius, $v+h$, and unit direction
  $\protect \bf n$ on the 3-sphere, constructed such that $\phi _I = (v+h)
  n_I$. Being a unit vector, we have ${\protect \bf n} \cdot {\protect \bf n} =
  1$, and it is parameterized by the three angular coordinates $\pi _i/v$, in
  detail $$ {\protect \bf n} = \begin {pmatrix} 0 \\ 0 \\ 0 \\ 1 \end {pmatrix}
  \protect \qopname \relax o{exp}[\pi _i \tau _i/v], $$ where again $\tau _i$
  are the generators of the rotations broken when the Higgs vev reduces
  custodial symmetry from ${\protect \rm O}(4)\rightarrow {\protect \rm O}(3)$.
  Using this language, the operators in the HEFT can be explicitly written in
  terms of ${\protect \bf n}$ rather than $\pi _i$. As a single example, we
  have $\protect \mathcal {O}_1^{h\pi } = v^2 (\partial h)^2(\partial _{\mu
  }{\protect \bf n}\cdot \partial ^{\mu }{\protect \bf n})$, up to terms of
  higher multiplicity in fields.

\bibitem{Note5}
More explicitly, the Wilson coefficients of SMEFT form a self-contained vector
  subspace within the space of HEFT coefficients. While a violation of the
  SMEFT bounds, and the observation of nonzero Wilson coefficients in the
  HEFT-allowed region of Fig.~\ref {fig:proj}, would arguably correspond to a
  smoking gun for the HEFT, the converse is not true; it is possible for the
  correct EFT to be the HEFT, but for its Wilson coefficients to be such that
  they do not allow the theory to be distinguished from the SMEFT via
  positivity bounds, e.g., if they land in the SMEFT-allowed region of
  Fig.~\ref {fig:proj}.

\bibitem{Salas-Bernardez:2022hqv}
A.~Salas-Bernardez, J.~J. Sanz-Cillero, F.~J. Llanes-Estrada, and
  R.~Gomez-Ambrosio, ``{SMEFT as a slice of HEFT\textquoteright{}s parameter
  space},'' \href{http://dx.doi.org/10.1051/epjconf/202227408013}{{\em EPJ Web
  Conf.} {\bfseries 274} (2022) 08013},
  \href{http://arxiv.org/abs/2211.09605}{{\ttfamily arXiv:2211.09605
  [hep-ph]}}.

\bibitem{Note6}
Throughout, we write all inequalities as strict for brevity. However, if we are
  working at fixed order in the coupling expansion in a weakly coupled theory,
  all of these results reduce to weak inequalities~\cite
  {Adams:2006sv,Nicolis:2009qm,Chandrasekaran:2018qmx,Remmen:2019cyz}.

\bibitem{Froissart:1961ux}
M.~Froissart, ``{Asymptotic behavior and subtractions in the Mandelstam
  representation},''
\href{http://dx.doi.org/10.1103/PhysRev.123.1053}{{\em Phys. Rev.} {\bfseries
  123} (1961) 1053}.
%%CITATION = PHRVA,123,1053;%%.

\bibitem{Martin:1962rt}
A.~Martin, ``{Unitarity and high-energy behavior of scattering amplitudes},''
\href{http://dx.doi.org/10.1103/PhysRev.129.1432}{{\em Phys. Rev.} {\bfseries
  129} (1963) 1432}.
%%CITATION = PHRVA,129,1432;%%.

\bibitem{Martin:1965jj}
A.~Martin, ``{Extension of the axiomatic analyticity domain of scattering
  amplitudes by unitarity. 1.},''
  \href{http://dx.doi.org/10.1007/BF02720568}{{\em Nuovo Cim. A} {\bfseries 42}
  (1965) 930}.

\bibitem{Mandelstam:1958xc}
S.~Mandelstam, ``{Determination of the Pion-Nucleon Scattering Amplitude from
  Dispersion Relations and Unitarity. General Theory},''
\href{http://dx.doi.org/10.1103/PhysRev.112.1344}{{\em Phys. Rev.} {\bfseries
  112} (1958) 1344}.
%%CITATION = PHRVA,112,1344;%%.

\bibitem{Lehmann:1958ita}
H.~Lehmann, ``{Analytic properties of scattering amplitudes as functions of
  momentum transfer},'' \href{http://dx.doi.org/10.1007/bf02859794}{{\em Nuovo
  Cim.} {\bfseries 10} (1958) 579}.

\bibitem{Note7}
See Refs.~\cite {Remmen:2019cyz,ZZ} for discussion of tree-level UV extensions
  of the SMEFT Higgs operators.

\bibitem{Note8}
That is, all of these interactions can be simply written as couplings of heavy
  states to the unit vector ${\protect \bf n}$. The only subtlety in writing
  the ${\protect \rm O}(4)$ analogue of our UV extensions is that for the $B$
  vector one first needs to dualize to an ${\protect \rm O}(4)$ valued
  two-form, i.e., $B^{ab}_{\mu }n^a\partial ^\mu n^b$.

\bibitem{Chala:2021wpj}
M.~Chala and J.~Santiago, ``{Positivity bounds in the standard model effective
  field theory beyond tree level},''
  \href{http://dx.doi.org/10.1103/PhysRevD.105.L111901}{{\em Phys. Rev. D}
  {\bfseries 105} no.~11, (2022) L111901},
  \href{http://arxiv.org/abs/2110.01624}{{\ttfamily arXiv:2110.01624
  [hep-ph]}}.

\bibitem{Chala:2021pll}
M.~Chala, G.~Guedes, M.~Ramos, and J.~Santiago, ``{Towards the renormalisation
  of the Standard Model effective field theory to dimension eight: Bosonic
  interactions I},''
  \href{http://dx.doi.org/10.21468/SciPostPhys.11.3.065}{{\em SciPost Phys.}
  {\bfseries 11} (2021) 065}, \href{http://arxiv.org/abs/2106.05291}{{\ttfamily
  arXiv:2106.05291 [hep-ph]}}.

\bibitem{Li:2022aby}
X.~Li, ``{Positivity bounds at one-loop level: the Higgs sector},''
  \href{http://dx.doi.org/10.1007/JHEP05(2023)230}{{\em JHEP} {\bfseries 05}
  (2023) 230}, \href{http://arxiv.org/abs/2212.12227}{{\ttfamily
  arXiv:2212.12227 [hep-ph]}}.

\bibitem{Liao:2025npz}
Y.-P. Liao, J.~Roosmale~Nepveu, and C.-H. Shen, ``{Positivity in Perturbative
  Renormalization: an EFT $a$-theorem},''
  \href{http://arxiv.org/abs/2505.02910}{{\ttfamily arXiv:2505.02910
  [hep-ph]}}.

\bibitem{Cappati:2022skp}
A.~Cappati, R.~Covarelli, P.~Torrielli, and M.~Zaro, ``{Sensitivity to new
  physics in final states with multiple gauge and Higgs bosons},''
  \href{http://dx.doi.org/10.1007/JHEP09(2022)038}{{\em JHEP} {\bfseries 09}
  (2022) 038}, \href{http://arxiv.org/abs/2205.15959}{{\ttfamily
  arXiv:2205.15959 [hep-ph]}}.

\bibitem{Henning:2015alf}
B.~Henning, X.~Lu, T.~Melia, and H.~Murayama, ``{$2,\, 84,\, 30,\, 993,\,
  560,\, 15456,\, 11962,\, 261485,\ldots\,$: Higher dimension operators in the
  SM EFT},'' \href{http://dx.doi.org/10.1007/JHEP08(2017)016}{{\em JHEP}
  {\bfseries 08} (2017) 016}, \href{http://arxiv.org/abs/1512.03433}{{\ttfamily
  arXiv:1512.03433 [hep-ph]}}. [Erratum:
  \href{https://doi.org/10.1007/JHEP09(2019)019}{{\it JHEP} {\bf 09} (2019)
  019}].

\bibitem{ATLAS:2018mxa}
{\bfseries ATLAS} {\bfseries Collaboration}, M.~Aaboud { et~al.},
  ``{Observation of electroweak $W^{\pm}Z$ boson pair production in association
  with two jets in $pp$ collisions at $\sqrt{s} =$ 13 TeV with the ATLAS
  detector},'' \href{http://dx.doi.org/10.1016/j.physletb.2019.05.012}{{\em
  Phys. Lett. B} {\bfseries 793} (2019) 469--492},
  \href{http://arxiv.org/abs/1812.09740}{{\ttfamily arXiv:1812.09740
  [hep-ex]}}.

\bibitem{ATLAS:2019thr}
{\bfseries ATLAS} {\bfseries Collaboration}, G.~Aad { et~al.}, ``{Search for
  the electroweak diboson production in association with a high-mass dijet
  system in semileptonic final states in $pp$ collisions at $\sqrt{s}=13$ TeV
  with the ATLAS detector},''
  \href{http://dx.doi.org/10.1103/PhysRevD.100.032007}{{\em Phys. Rev. D}
  {\bfseries 100} (2019) 032007},
  \href{http://arxiv.org/abs/1905.07714}{{\ttfamily arXiv:1905.07714
  [hep-ex]}}.

\bibitem{ATLAS:2019qhm}
{\bfseries ATLAS} {\bfseries Collaboration}, G.~Aad { et~al.}, ``{Evidence for
  electroweak production of two jets in association with a $Z\gamma$ pair in
  $pp$ collisions at $\sqrt{s} = 13$ TeV with the ATLAS detector},''
  \href{http://dx.doi.org/10.1016/j.physletb.2020.135341}{{\em Phys. Lett. B}
  {\bfseries 803} (2020) 135341},
  \href{http://arxiv.org/abs/1910.09503}{{\ttfamily arXiv:1910.09503
  [hep-ex]}}.

\bibitem{ATLAS:2020nlt}
{\bfseries ATLAS} {\bfseries Collaboration}, G.~Aad { et~al.}, ``{Observation
  of electroweak production of two jets and a Z-boson pair},''
  \href{http://dx.doi.org/10.1038/s41567-022-01757-y}{{\em Nature Phys.}
  {\bfseries 19} (2023) 237}, \href{http://arxiv.org/abs/2004.10612}{{\ttfamily
  arXiv:2004.10612 [hep-ex]}}.

\bibitem{ATLAS:2024ini}
{\bfseries ATLAS} {\bfseries Collaboration}, G.~Aad { et~al.}, ``{Measurements
  of electroweak W$^{±}$Z boson pair production in association with two jets
  in pp collisions at $ \sqrt{s} $ = 13 TeV with the ATLAS detector},''
  \href{http://dx.doi.org/10.1007/JHEP06(2024)192}{{\em JHEP} {\bfseries 06}
  (2024) 192}, \href{http://arxiv.org/abs/2403.15296}{{\ttfamily
  arXiv:2403.15296 [hep-ex]}}.

\bibitem{CMS:2019uys}
{\bfseries CMS} {\bfseries Collaboration}, A.~M. Sirunyan { et~al.},
  ``{Measurement of electroweak WZ boson production and search for new physics
  in WZ + two jets events in pp collisions at $\sqrt{s} =$ 13 TeV},''
  \href{http://dx.doi.org/10.1016/j.physletb.2019.05.042}{{\em Phys. Lett. B}
  {\bfseries 795} (2019) 281},
  \href{http://arxiv.org/abs/1901.04060}{{\ttfamily arXiv:1901.04060
  [hep-ex]}}.

\bibitem{CMS:2019mpq}
{\bfseries CMS} {\bfseries Collaboration}, A.~M. Sirunyan { et~al.}, ``{Search
  for the production of W$^\pm$W$^\pm$W$^\mp$ events at $\sqrt{s} =$ 13 TeV},''
  \href{http://dx.doi.org/10.1103/PhysRevD.100.012004}{{\em Phys. Rev. D}
  {\bfseries 100} (2019) 012004},
  \href{http://arxiv.org/abs/1905.04246}{{\ttfamily arXiv:1905.04246
  [hep-ex]}}.

\bibitem{CMS:2019qfk}
{\bfseries CMS} {\bfseries Collaboration}, A.~M. Sirunyan { et~al.}, ``{Search
  for anomalous electroweak production of vector boson pairs in association
  with two jets in proton-proton collisions at 13 TeV},''
  \href{http://dx.doi.org/10.1016/j.physletb.2019.134985}{{\em Phys. Lett. B}
  {\bfseries 798} (2019) 134985},
  \href{http://arxiv.org/abs/1905.07445}{{\ttfamily arXiv:1905.07445
  [hep-ex]}}.

\bibitem{CMS:2020ioi}
{\bfseries CMS} {\bfseries Collaboration}, A.~M. Sirunyan { et~al.},
  ``{Measurement of the cross section for electroweak production of a Z boson,
  a photon and two jets in proton-proton collisions at $\sqrt{s} =$ 13 TeV and
  constraints on anomalous quartic couplings},''
  \href{http://dx.doi.org/10.1007/JHEP06(2020)076}{{\em JHEP} {\bfseries 06}
  (2020) 076}, \href{http://arxiv.org/abs/2002.09902}{{\ttfamily
  arXiv:2002.09902 [hep-ex]}}.

\bibitem{CMS:2020gfh}
{\bfseries CMS} {\bfseries Collaboration}, A.~M. Sirunyan { et~al.},
  ``{Measurements of production cross sections of WZ and same-sign WW boson
  pairs in association with two jets in proton-proton collisions at $\sqrt{s}
  =$ 13 TeV},'' \href{http://dx.doi.org/10.1016/j.physletb.2020.135710}{{\em
  Phys. Lett. B} {\bfseries 809} (2020) 135710},
  \href{http://arxiv.org/abs/2005.01173}{{\ttfamily arXiv:2005.01173
  [hep-ex]}}.

\bibitem{CMS:2020fqz}
{\bfseries CMS} {\bfseries Collaboration}, A.~M. Sirunyan { et~al.},
  ``{Evidence for electroweak production of four charged leptons and two jets
  in proton-proton collisions at $\sqrt {s}$ = 13 TeV},''
  \href{http://dx.doi.org/10.1016/j.physletb.2020.135992}{{\em Phys. Lett. B}
  {\bfseries 812} (2021) 135992},
  \href{http://arxiv.org/abs/2008.07013}{{\ttfamily arXiv:2008.07013
  [hep-ex]}}.

\bibitem{CMS:2020ypo}
{\bfseries CMS} {\bfseries Collaboration}, A.~M. Sirunyan { et~al.},
  ``{Observation of electroweak production of W$\gamma$ with two jets in
  proton-proton collisions at $\sqrt {s}$ = 13 TeV},''
  \href{http://dx.doi.org/10.1016/j.physletb.2020.135988}{{\em Phys. Lett. B}
  {\bfseries 811} (2020) 135988},
  \href{http://arxiv.org/abs/2008.10521}{{\ttfamily arXiv:2008.10521
  [hep-ex]}}.

\bibitem{CMS:2021jji}
{\bfseries CMS} {\bfseries Collaboration}, A.~Tumasyan { et~al.},
  ``{Measurements of the pp $\to$ W$^\pm\gamma\gamma$ and pp $\to$
  Z$\gamma\gamma$ cross sections at $\sqrt{s} =$ 13 TeV and limits on anomalous
  quartic gauge couplings},''
  \href{http://dx.doi.org/10.1007/JHEP10(2021)174}{{\em JHEP} {\bfseries 10}
  (2021) 174}, \href{http://arxiv.org/abs/2105.12780}{{\ttfamily
  arXiv:2105.12780 [hep-ex]}}.

\bibitem{CMS:2021gme}
{\bfseries CMS} {\bfseries Collaboration}, A.~Tumasyan { et~al.},
  ``{Measurement of the electroweak production of Z$\gamma$ and two jets in
  proton-proton collisions at $\sqrt{s} =$ 13 TeV and constraints on anomalous
  quartic gauge couplings},''
  \href{http://dx.doi.org/10.1103/PhysRevD.104.072001}{{\em Phys. Rev. D}
  {\bfseries 104} (2021) 072001},
  \href{http://arxiv.org/abs/2106.11082}{{\ttfamily arXiv:2106.11082
  [hep-ex]}}.

\bibitem{CMS:2022dmc}
{\bfseries CMS, TOTEM} {\bfseries Collaboration}, A.~Tumasyan { et~al.},
  ``{Search for high-mass exclusive $\gamma\gamma\to WW$ and $\gamma\gamma\to
  ZZ$ production in proton-proton collisions at $\sqrt{s}$ = 13 TeV},''
  \href{http://dx.doi.org/10.1007/JHEP07(2023)229}{{\em JHEP} {\bfseries 07}
  (2023) 229}, \href{http://arxiv.org/abs/2211.16320}{{\ttfamily
  arXiv:2211.16320 [hep-ex]}}.

\bibitem{CMS:2022yrl}
{\bfseries CMS} {\bfseries Collaboration}, A.~Tumasyan { et~al.},
  ``{Measurement of the electroweak production of W$\gamma$ in association with
  two jets in proton-proton collisions at $\sqrt{s}$ = 13 TeV},''
  \href{http://dx.doi.org/10.1103/PhysRevD.108.032017}{{\em Phys. Rev. D}
  {\bfseries 108} (2023) 032017},
  \href{http://arxiv.org/abs/2212.12592}{{\ttfamily arXiv:2212.12592
  [hep-ex]}}.

\bibitem{Bellazzini:2015cra}
B.~Bellazzini, C.~Cheung, and G.~N. Remmen, ``{Quantum Gravity Constraints from
  Unitarity and Analyticity},''
  \href{http://dx.doi.org/10.1103/PhysRevD.93.064076}{{\em Phys. Rev. D}
  {\bfseries 93} (2016) 064076},
  \href{http://arxiv.org/abs/1509.00851}{{\ttfamily arXiv:1509.00851
  [hep-th]}}.

\bibitem{Cheung:2016wjt}
C.~Cheung and G.~N. Remmen, ``{Positivity of Curvature-Squared Corrections in
  Gravity},'' \href{http://dx.doi.org/10.1103/PhysRevLett.118.051601}{{\em
  Phys. Rev. Lett.} {\bfseries 118} (2017) 051601},
  \href{http://arxiv.org/abs/1608.02942}{{\ttfamily arXiv:1608.02942
  [hep-th]}}.

\bibitem{Camanho:2014apa}
X.~O. Camanho, J.~D. Edelstein, J.~Maldacena, and A.~Zhiboedov, ``{Causality
  Constraints on Corrections to the Graviton Three-Point Coupling},''
  \href{http://dx.doi.org/10.1007/JHEP02(2016)020}{{\em JHEP} {\bfseries 02}
  (2016) 020}, \href{http://arxiv.org/abs/1407.5597}{{\ttfamily arXiv:1407.5597
  [hep-th]}}.

\bibitem{Gruzinov:2006ie}
A.~Gruzinov and M.~Kleban, ``{Causality Constrains Higher Curvature Corrections
  to Gravity},'' \href{http://dx.doi.org/10.1088/0264-9381/24/13/N02}{{\em
  Class. Quant. Grav.} {\bfseries 24} (2007) 3521},
  \href{http://arxiv.org/abs/hep-th/0612015}{{\ttfamily arXiv:hep-th/0612015}}.

\bibitem{Cheung:2014ega}
C.~Cheung and G.~N. Remmen, ``{Infrared Consistency and the Weak Gravity
  Conjecture},'' \href{http://dx.doi.org/10.1007/JHEP12(2014)087}{{\em JHEP}
  {\bfseries 12} (2014) 087},
\href{http://arxiv.org/abs/1407.7865}{{\ttfamily arXiv:1407.7865 [hep-th]}}.
%%CITATION = ARXIV:1407.7865;%%.

\bibitem{Bellazzini:2019xts}
B.~Bellazzini, M.~Lewandowski, and J.~Serra, ``{Amplitudes' Positivity, Weak
  Gravity Conjecture, and Modified Gravity},''
\href{http://arxiv.org/abs/1902.03250}{{\ttfamily arXiv:1902.03250 [hep-th]}}.
%%CITATION = ARXIV:1902.03250;%%.

\bibitem{Andriolo:2020lul}
S.~Andriolo, T.-C. Huang, T.~Noumi, H.~Ooguri, and G.~Shiu, ``{Duality and
  axionic weak gravity},''
  \href{http://dx.doi.org/10.1103/PhysRevD.102.046008}{{\em Phys. Rev. D}
  {\bfseries 102} (2020) 046008},
  \href{http://arxiv.org/abs/2004.13721}{{\ttfamily arXiv:2004.13721
  [hep-th]}}.

\bibitem{Caron-Huot:2021rmr}
S.~Caron-Huot, D.~Mazac, L.~Rastelli, and D.~Simmons-Duffin, ``{Sharp
  boundaries for the swampland},''
  \href{http://dx.doi.org/10.1007/JHEP07(2021)110}{{\em JHEP} {\bfseries 07}
  (2021) 110}, \href{http://arxiv.org/abs/2102.08951}{{\ttfamily
  arXiv:2102.08951 [hep-th]}}.

\bibitem{Caron-Huot:2022ugt}
S.~Caron-Huot, Y.-Z. Li, J.~Parra-Martinez, and D.~Simmons-Duffin, ``{Causality
  constraints on corrections to Einstein gravity},''
  \href{http://dx.doi.org/10.1007/JHEP05(2023)122}{{\em JHEP} {\bfseries 05}
  (2023) 122}, \href{http://arxiv.org/abs/2201.06602}{{\ttfamily
  arXiv:2201.06602 [hep-th]}}.

\bibitem{Caron-Huot:2022jli}
S.~Caron-Huot, Y.-Z. Li, J.~Parra-Martinez, and D.~Simmons-Duffin, ``{Graviton
  partial waves and causality in higher dimensions},''
  \href{http://dx.doi.org/10.1103/PhysRevD.108.026007}{{\em Phys. Rev. D}
  {\bfseries 108} (2023) 026007},
  \href{http://arxiv.org/abs/2205.01495}{{\ttfamily arXiv:2205.01495
  [hep-th]}}.

\bibitem{Cheung:2016yqr}
C.~Cheung and G.~N. Remmen, ``{Positive Signs in Massive Gravity},''
  \href{http://dx.doi.org/10.1007/JHEP04(2016)002}{{\em JHEP} {\bfseries 04}
  (2016) 002},
\href{http://arxiv.org/abs/1601.04068}{{\ttfamily arXiv:1601.04068 [hep-th]}}.
%%CITATION = ARXIV:1601.04068;%%.

\bibitem{deRham:2017xox}
C.~de~Rham, S.~Melville, and A.~J. Tolley, ``{Improved Positivity Bounds and
  Massive Gravity},'' \href{http://dx.doi.org/10.1007/JHEP04(2018)083}{{\em
  JHEP} {\bfseries 04} (2018) 083},
\href{http://arxiv.org/abs/1710.09611}{{\ttfamily arXiv:1710.09611 [hep-th]}}.
%%CITATION = ARXIV:1710.09611;%%.

\bibitem{Camanho:2016opx}
X.~O. Camanho, G.~Lucena~G\'omez, and R.~Rahman, ``{Causality Constraints on
  Massive Gravity},'' \href{http://dx.doi.org/10.1103/PhysRevD.96.084007}{{\em
  Phys. Rev. D} {\bfseries 96} (2017) 084007},
  \href{http://arxiv.org/abs/1610.02033}{{\ttfamily arXiv:1610.02033
  [hep-th]}}.

\bibitem{Chandrasekaran:2018qmx}
V.~Chandrasekaran, G.~N. Remmen, and A.~Shahbazi-Moghaddam, ``{Higher-Point
  Positivity},'' \href{http://dx.doi.org/10.1007/JHEP11(2018)015}{{\em JHEP}
  {\bfseries 11} (2018) 015}, \href{http://arxiv.org/abs/1804.03153}{{\ttfamily
  arXiv:1804.03153 [hep-th]}}.

\bibitem{Kallen}
G.~K{\"a}ll{\'e}n, ``{On the Definition of the Renormalization Constants in
  Quantum Electrodynamics},''
  \href{http://dx.doi.org/10.1007/978-3-319-00627-7_90}{{\em Helv. Phys. Acta}
  {\bfseries 25} (1952) 417}.

\bibitem{Lehmann}
H.~Lehmann, ``{On the Properties of Propagation Functions and Renormalization
  Contants of Quantized Fields},''
\href{http://dx.doi.org/10.1007/BF02783624}{{\em Nuovo Cim.} {\bfseries 11}
  (1954) 342}.
%%CITATION = NUCIA,11,342;%%.

\bibitem{Cutkosky:1960sp}
R.~E. Cutkosky, ``{Singularities and discontinuities of Feynman amplitudes},''
  \href{http://dx.doi.org/10.1063/1.1703676}{{\em J. Math. Phys.} {\bfseries 1}
  (1960) 429}.

\bibitem{Note9}
See the discussion in Sec. 2.3.1 of Ref.~\cite {Trott}.

\end{thebibliography}\endgroup

\clearpage
\appendix
\onecolumngrid
\section*{End Matter}
\twocolumngrid

\setcounter{equation}{0}
\renewcommand{\theequation}{A\arabic{equation}}

%%%%%%%%%%%%%%%%%%%%%%%%%%%%%%%%%
\mysec{Appendix A: All Possible Superpositions}
%%%%%%%%%%%%%%%%%%%%%%%%%%%%%%%%%
%
Starting from the expression for $A''(s)$ in \Eq{eq:ddA}, let us see how the positivity bound $A''(s)>0$ from the optical theorem reduces to the set of conditions in \Eq{eq:bounds} after marginalizing over all possible superpositions of scattering states.
Let us define parameters $\mu,\nu\geq 0$ and $\rho \in [-1,1]$ via $\mu=\alpha/(1-\alpha^{2})^{1/2}$, $\nu=\beta/(1-\beta^{2})^{1/2}$, and $\rho=\boldsymbol{\alpha}\cdot \boldsymbol{\beta}/\alpha\beta$,
as well as a reparameterization of the Wilson coefficients,
\bea
x & =\frac{\cHb}{\sqrt{4\cHa(\cHd+\cHe)}},&
y & =\frac{\cHc}{\sqrt{4\cHa(\cHd+\cHe)}},
\\
z & =\sqrt{\frac{\cHa}{\cHd+\cHe}}, &
w & =\sqrt{\frac{\cHe}{\cHd+\cHe}}.
\label{eq:xyzw}
\eea
The bounds on the Wilson coefficients then become the requirement that $(x,y,z,w)$ satisfy
\bea
0<&\,z^{2}+\mu^{2}\nu^{2}\rho^{2}+\frac{1}{2}yz(\mu^{2}+2\mu\nu \rho+\nu^{2})
\\
&+2xz\mu\nu \rho +\frac{1}{2}w^{2}\mu^{2}\nu^{2}(1-\rho^{2})
\label{eq:boundxyz}
\eea
for all $\mu,\nu\geq 0$ and $-1\leq \rho \leq 1$.
Imposing $A''(s)>0$ on \Eq{eq:ddA} for $\alpha=\beta=0$, we have $\cHa>0$, while from $\alpha=\beta=1$ and ${\boldsymbol{\alpha}} = {\boldsymbol{\beta}}$, we have $\cHd+\cHe>0$.
Meanwhile, with $\alpha=\beta=1$ and choosing ${\boldsymbol{\alpha}}\perp{\boldsymbol{\beta}}$, we find $\cHe>0$, while with $\beta=1$ and $\alpha=0$, we obtain $\cHc>0$.
In terms of \Eq{eq:xyzw}, we therefore have
\be 
y,z,w>0\hspace{0.2cm}
{\rm and}\hspace{0.2cm}
x\in \mathbb{R}.
\label{eq:prereq}
\ee

First marginalizing \Eq{eq:boundxyz} over all $\mu,\nu\geq 0$, we find the condition
\be
y+(2x+y)\rho + \sqrt{4\rho^2+2w^2(1-\rho^2)}>0.
\label{eq:rhocond}
\ee
Marginalizing \Eq{eq:rhocond} over all $\rho\in [-1,1]$ while imposing \Eq{eq:prereq}, we arrive at the conditions
\be
y,z,w>0,\hspace{0.2cm}
x<1,\hspace{0.2cm}
{\rm and}\hspace{0.2cm}
1+x+y>0.
\label{eq:xycond}
\ee
Putting the definitions in \Eq{eq:xyzw} together with \Eq{eq:xycond} and the positivity bounds in the previous paragraph, we obtain the final set of conditions on the Wilson coefficients given in \Eq{eq:bounds}.

%%%%%%%%%%%%%%%%%%%%%%%%%%%%%%%%%
\medskip
\setcounter{equation}{0}
\renewcommand{\theequation}{B\arabic{equation}}
\mysec{Appendix B: Generalized Optical Theorem}
%%%%%%%%%%%%%%%%%%%%%%%%%%%%%%%%%
%
Elastic positivity bounds result from the standard optical theorem, ${\rm Im}\,A(s) = s\,\sigma(s)$, where $A$ is a forward amplitude with the two-particle ingoing state matching the outgoing one.
However, this is a special case of the {\it generalized} optical theorem, which allows us to consider cases where the two states are not identical.
Specifically, scattering $\Phi_I\Phi_J\rightarrow\Phi_K\Phi_L$ with zero momentum transfer ($t\,{=}\,0$) and center-of-mass energy squared $s$, described by an amplitude $A_{IJKL}(s)$, the generalized optical theorem states that ${\rm Disc}\, A_{IJKL}(s) = i\sum_X M_{IJ\rightarrow X}(s)M^*_{KL\rightarrow X}(s)$, where $M_{IJ\rightarrow X}$ is the amplitude for $\Phi_I\Phi_J\rightarrow X$ for any intermediate state $X$, and $\sum_X$ denotes the K\"all\'en-Lehmann-like sum over all such states~\cite{Kallen,Lehmann}, including multi-particle loops treated as integrals over on-shell configurations~\cite{Arkani-Hamed:2020blm,Cutkosky:1960sp}.
From a dispersion relation construction analogous to Eq.~\eqref{eq:dispersion} for $A''_{IJKL} = \lim_{s\rightarrow 0} \partial_s^2 A_{IJKL}(s)$, one finds~\cite{Arkani-Hamed:2021ajd,Freytsis:2022aho} that $A''_{IJKL}$ in Eq.~\eqref{eq:ddApre} must satisfy the relation
\be
A''_{IJKL} = \sum_X \left(M^{(X)}_{IJ} M^{(X)}_{KL} + M^{(X)}_{IL}M^{(X)}_{JK}\right)
\label{eq:mm}\vspace{-4mm}
\ee
for some collection of real matrices $M^{(X)}_{IJ}$ that run over the real and imaginary parts of $M_{IJ\rightarrow X}$.
Finding the space of Wilson coefficients parameterized by $A''_{IJKL}$ and swept out by all choices of $M^{(X)}_{IJ}$ is equivalent to marginalizing over all UV completions of the EFT and is in general a highly complex problem, which can nonetheless be solved explicitly in certain special cases~\cite{Freytsis:2022aho}.

However, we can decompose the right-hand side of Eq.~\eqref{eq:mm} into irreducible representations of the symmetry group, and in doing so we find a set of candidate so-called ``extremal rays''~\cite{ZZ}.
One can show that the cone of allowed $A''_{IJKL}$ is spanned by the convex hull of all such extremal rays.
In physical terms, these rays correspond to one-particle UV extensions of the EFT operators.
In the case at hand, each of our external states comprises a multiplet $\Phi_I = (\pi_i,h)$, with $\pi_i$ and $h$ transforming as a vector and singlet of the ${\rm O}(3)$ custodial symmetry, respectively.
Hence, we must construct the irreducible representations of $\mathbf{R}\otimes \mathbf{R}$, where $\mathbf{R}=\mathbf{3}\oplus\mathbf{1}$.
Our analysis here complements that for elastic bounds in Eq.~\eqref{eq:bounds}, which were necessary but in principle not sufficient for unitarity.
We find that the bounds constructed with these two methods precisely match, so the constraints in Eq.~\eqref{eq:bounds} are indeed both necessary and sufficient.
Expanding the product, we have $\mathbf{R}\otimes\mathbf{R} = (\mathbf{3}\otimes\mathbf{3}) \oplus (\mathbf{3}\otimes\mathbf{1}) \oplus (\mathbf{1}\otimes\mathbf{3}) \oplus (\mathbf{1}\otimes\mathbf{1})$.
The final three parenthetical terms are already irreducible to two vectors and a singlet.
Viewing the intermediate state $X$ in $\Phi_I \Phi_J \rightarrow X_{IJ}$ as a matrix,
\be 
X=\begin{pmatrix}
X_{ij} & X_{i4}\\
X_{4i} & X_{44} 
\end{pmatrix}\!,
\ee
$X_{44}$ is the singlet $\mathbf{1}\otimes\mathbf{1}=\mathbf{1}$, $X_{4i}$ and $X_{i4}$ are the vectors $\mathbf{1}\otimes\mathbf{3}=\mathbf{3}$ and $\mathbf{3}\otimes\mathbf{1}=\mathbf{3}$, and the $\mathbf{3}\otimes\mathbf{3}=\mathbf{9}$ term, corresponding to the matrix $X_{ij}$, decomposes as $\mathbf{9}=\mathbf{1}\oplus\mathbf{3}\oplus\mathbf{5}$, where $\mathbf{1}$ corresponds to the trace, $\mathbf{3}$ to the antisymmetric part (the dual of a vector), and $\mathbf{5}$ to the symmetric traceless matrices.

If there is no degeneracy in the group structure of our exchanged states---i.e., if we only have at most one of each type of irreducible representation---then we can construct projectors onto each of those states, and these projectors define candidate extremal rays that form the cone giving us the optimal bounds.
However, in the case above, we have two singlets in $\mathbf{R}\otimes\mathbf{R}$: one from the product of the $\mathbf{1}$ factors in $\mathbf{R}$, and the $\mathbf{1}$ in the decomposition of the $\mathbf{9}$.
In more physical terms, if we scatter two $\pi_{i}$ states, they can form a state in the UV that is a singlet under custodial symmetry, which could then decay to two $h$ particles, or alternatively into two $\pi_{i}$. Similarly, we have three triplets: the $X_{4i}$ and $X_{i4}$ components, as well as the $\mathbf{3}$ from the $\mathbf{9}$.
We only lack degeneracy in the $\mathbf{5}$.
When there is degeneracy, the extremal cone acquires curved facets, associated with the continuous families of candidate extremal rays parameterized by the couplings allowing the degenerate states to transition into each other~\footnote{See the discussion in Sec. 2.3.1 of Ref.~\cite{Trott}.}.

We are interested in constructing the projection operators $P_{IJKL}$ onto the above irreducible representations. That is, for an arbitrary matrix $X_{KL}$, $P_{IJKL}X_{KL}$ is the projection onto the components of the desired representation.
In the presence of degeneracy, the $P$ tensors contain free parameters in order to account for the freedom of the couplings mixing the representations.
Conveniently for our present purposes, the system of $W$ and $B$ fields in the SM has a group theoretic structure that maps neatly onto $\mathbf{R}$: there is a singlet $B$ boson and an ${\rm SU}(2)$ triplet of $W$ bosons (which, up to the $\mathbb{Z}_2$ of $\pi_i\leftrightarrow -\pi_i$, is locally equivalent to our ${\rm O}(3)$ custodial group).
The projectors for this system were determined explicitly in Sec.~5 of Ref.~\cite{YZZ}.
We write $\bar\delta_{IJ}(r)=\delta_{IJ}+(r-1)\delta_{I4}\delta_{J4}$, so that the $\bar\delta_{IJ}$ defined in the main text is $\bar\delta_{IJ}(0)$.
The projector onto the singlet is
\be 
P^{\bf 1}_{IJKL}(r)=\frac{1}{3}\bar\delta_{IJ}(r)\bar\delta_{KL}(r).
\ee
For the triplets, we first define the following matrices,
\bea
f^{(1)}_{IJ}(r_{1},r_{2}) & =\left(\begin{smallmatrix}
0 & 0 & 0 & r_{1}\\
0 & 0 & 1 & 0\\
0 & -1 & 0 & 0\\
r_{2} & 0 & 0 & 0
\end{smallmatrix}\right)\\
f^{(2)}_{IJ}(r_{1},r_{2}) & =\left(\begin{smallmatrix}
0 & 0 & -1 & 0\\
0 & 0 & 0 & r_{1}\\
1 & 0 & 0 & 0\\
0 & r_{2} & 0 & 0
\end{smallmatrix}\right)\\
f^{(3)}_{IJ}(r_{1},r_{2}) & =\left(\begin{smallmatrix}
0 & 1 & 0 & 0\\
-1 & 0 & 0 & 0\\
0 & 0 & 0 & r_{1}\\
0 & 0 & r_{2} & 0
\end{smallmatrix}\right)\!.
\eea
As one can explicitly verify, we have two projectors onto the triplet states,
\bea
P^{{\bf 3}+}_{IJKL}(r_{1},r_{2})&=\frac{1}{2}\sum_{i=0}^{3}f^{(i)}_{(IJ)}(r_{1},r_{2})f^{(i)}_{(KL)}(r_{1},r_{2})\\
P^{{\bf 3}-}_{IJKL}(r_{1},r_{2})&=\frac{1}{2}\sum_{i=0}^{3}f^{(i)}_{[IJ]}(r_{1},r_{2})f^{(i)}_{[KL]}(r_{1},r_{2}).
\eea
Noting that $\bar\delta_{IJ}(0)$ defines a projector onto the subspace where $I,J\neq4$, we can define the projector of the $\mathbf{5}$,
\be
P^{{\bf 5}}_{IJKL}=\bar\delta_{I(K}(0)\bar\delta_{L)J}(0)-\frac{1}{3}\bar\delta_{IJ}(0)\bar\delta_{KL}(0).
\ee

\newpage
At last, we define the candidate extremal rays by symmetrizing on $J\leftrightarrow L$, which we write as $\hat{P}_{IJKL}=(P_{IJKL}+P_{ILKJ})/2$.
We find that $\hat{P}^{{\bf 3}+}_{IJKL}$ is proportional to $(r_1+r_2)^{2}$.
Since we are concerned with rays, we can divide out by this combination in $\hat P^{{\bf 3}+}_{IJKL}$.
Similarly, $\hat{P}^{{\bf 3}-}_{IJKL}$ depends only on $r_1-r_2$, which we will denote by $q$.
We find that $\hat P^{{\bf 5}}_{IJKL}$ is redundant, since it can be written as a positive linear combination of the others, $\hat{P}^{\bf 5}_{IJKL}=2\hat{P}^{\bf 1}_{IJKL}(0)+\hat{P}^{{\bf 3}-}_{IJKL}(0)$, so we discard it.
Thus, our candidate extremal rays are
\be 
\hat{P}^{\bf 1}_{IJKL}(r),\;\;\hat{P}^{{\bf 3}-}_{IJKL}(q),\;\;{\rm and}\;\;\hat{P}^{{\bf 3}+}_{IJKL}.
\ee
Two depend on parameters, and one does not.

The power of the generalized optical theorem is that we can replace Eq.~\eqref{eq:mm} with $A''_{IJKL} = \sum_\alpha N_\alpha \hat P^\alpha_{IJKL}$, where $\alpha$ runs over all of the labels of the projectors, including an integral over all possible parameters $q$ and $r$.
Matching with $A''_{IJKL} = 4(c_{IJKL} + c_{ILKJ})$ from Eq.~\eqref{eq:ddApre}, where the Wilson coefficient tensor is given in Eq.~\eqref{eq:Wilson} in terms of the HEFT coefficients defined in Eq.~\eqref{eq:HEFT}, we find
\bea
\int_{-\infty}^\infty {\rm d}r\, N_{\bf 1}(r) & =24(\cHd+\cHe)\\
\int_{-\infty}^\infty {\rm d}q\, N_{{\bf 3}-}(q) & =8\cHe\\
3N_{{\bf 3}+}+4\int_{-\infty}^\infty {\rm d}r\,r N_{\bf 1}(r) & =48(\cHb+\cHc)\\
N_{{\bf 3}+}+\int_{-\infty}^\infty {\rm d}q\,q^2 N_{{\bf 3}-}(q) & =16\cHc\\
\int_{-\infty}^\infty {\rm d}r\,r^2 N_{\bf 1}(r) & =24\cHa.
\label{eq:distribution}
\eea
Unitarity bounds the HEFT coefficients to be such that there exists a positive constant $N_{{\bf 3}+}$ and positive functions $N_{\bf 1}(r)$ and $N_{{\bf 3}-}(q)$ such that the constraints in Eq.~\eqref{eq:distribution} are satisfied.
To turn this statement into a bound on the HEFT coefficients alone, let us first suppose that $N_{\bf 1}(r) = n_{\bf 1}\delta(r-r_0)$ and $N_{{\bf 3}-}(q)=n_{{\bf 3}-}\delta(q-q_0)$.
Then we require that there exist some real values of $r_0$ and $q_0$ and some positive $n_{\bf 1}$, $n_{{\bf 3}-}$, and $N_{{\bf 3}+}$ such that Eq.~\eqref{eq:distribution} is satisfied.
This marginalization can be done in closed form, and in terms of Eq.~\eqref{eq:xyzw} we find
\be
y,z,w\,{>}\,0\,\,{\rm and}\,\,\bigl(x\,{<}-1\,{<}\,x{+}y\,\,{\rm or}\,\, x\,{<}\,1\,{<}\,x{+}y\bigr).
\label{eq:prehull}
\ee
In Eq.~\eqref{eq:prehull}, we required fixed values of $q$ and $r$. In actuality, each value of $q$ and $r$ defines a new extremal ray. To find the true bounds, we must therefore take the convex hull over the bounds implied by all possible choices of $(q,r)$.
The convex hull in $(x,y)$ from Eq.~\eqref{eq:prehull} gives precisely the set of conditions in Eq.~\eqref{eq:xycond} that we found from the elastic scattering dispersion relation.
Thus, the bounds on the HEFT that we obtain from the generalized optical theorem are precisely those in Eq.~\eqref{eq:bounds}, giving the necessary and sufficient conditions for unitarity.

\end{document}